\documentstyle[aps,pra,graphicx,floats]{revtex}

\begin{document}
\draft
\title{Theory of Pseudomodes in Quantum Optical Processes}
\author{B.~J.~Dalton$^{1,2}$, S.~M.~Barnett$^{3}$, and B.~M.~Garraway$^{1}$}
\address{$^{1}$Sussex Centre for Optical and Atomic Physics, University of Sussex,\\
Brighton BN1 9QH, United Kingdom\\
$^{2}$Department of Physics, University of Queensland, St Lucia, Queensland\\
4072, Australia\\
$^{3}$Department of Physics and Applied Physics, University of Strathclyde,\\
Glasgow G4 0NG, United Kingdom}
\date{February 22, 2001}
\maketitle

\begin{abstract}
\quad This paper deals with non-Markovian behaviour in atomic systems
coupled to a structured reservoir of quantum EM\ field modes, with
particular relevance to atoms interacting with the field in high Q cavities
or photonic band gap materials. In cases such as the former, we show that
the pseudo mode theory for single quantum reservoir excitations can be
obtained by applying the Fano diagonalisation method to a system in which
the atomic transitions are coupled to a discrete set of (cavity) quasimodes,
which in turn are coupled to a continuum set of (external) quasimodes with
slowly varying coupling constants and continuum mode density. Each
pseudomode can be identified with a discrete quasimode, which gives
structure to the actual reservoir of true modes via the expressions for the
equivalent atom-true mode coupling constants. The quasimode theory enables
cases of multiple excitation of the reservoir to now be treated via
Markovian master equations for the atom-discrete quasimode system.
Applications of the theory to one, two and many discrete quasimodes are
made. For a simple photonic band gap model, where the reservoir structure is
associated with the true mode density rather than the coupling constants,
the single quantum excitation case appears to be equivalent to a case with
two discrete quasimodes.
\end{abstract}


\section{ Introduction}

\label{SIntro}

The quantum behaviour of a small system coupled to a large one has been the
subject of many studies since quantum theory was first formulated. The small
system is usually of primary interest and generally microscopic (atom,
nucleus, molecule or small collections of these) but currently systems of a
more macroscopic nature (Bose condensate, superconductor, quantum computer)
are being studied. The large system is invariably macroscopic in nature
(free space or universe modes of the EM field, lattice modes in a solid,
collider atoms in a gas) and is of less interest in its own right, being
primarily of relevance as a reservoir or bath affecting the small system in
terms of relaxation and noise processes. The large system is often a model
for the entire external environment surrounding the small system. Changes in
the small system states (described in terms of its density operator) can be
divided into two sorts---effects on the state populations (energy loss or
gain) or effects on the state coherences (decoherence or induced coherence).
Equivalently, quantum information (described via the von Neumann entropy)
would be lost or gained due to the interaction with the environment, and its
loss is generally associated with decoherence. Interestingly, as the small
system becomes larger or occupies states that are more classical the time
scale for decoherence can become much smaller than that for energy loss.
This is of special interest in quantum information processing \cite
{Nielsen2000,Vedral98,Steane97} where the small system is a collection of
qbits making up a quantum computer weakly coupled to the outside world, or
in measurement theory \cite{Zurek91,Zurek98,Peres93,Isham95}, where the
small system is a micro system being measured coupled to an apparatus (or
pointer) that registers the results. For quantum computers it is desirable
that decoherence is negligible during the overall computation time \cite
{DiVincenzo2000} (otherwise error correction methods have to be
incorporated, and this is costly in terms of processing time), whereas in
measurement theory environment induced decoherence \cite{Zurek91,Haroche98}
is responsible for the density operator becoming diagonal in the pointer
basis (otherwise a macroscopic superposition of pointer readings would
result).

A standard method for describing the reservoir effects on the small system
is based on the Born-Markoff master equation for the system density operator 
\cite{Walls94,Scully97,Barnett97,Gardiner91}. This depends on the
correlation time for the reservoir (as determined from the behaviour of two
time correlation functions for pairs of reservoir operators involved in the
system reservoir interaction) being very short compared to that of the
relaxation and noise processes of the system. In general terms, the more
slowly varying the coupling constants for this interaction or the density of
reservoir states are with reservoir frequencies, the shorter the correlation
time will be. For many situations in the fields of quantum optics, NMR,
solid state physics the Born-Markoff master equation provided an accurate
description of the physics for the system. Elaborations or variants of the
method such as quantum state trajectories \cite{Carmichael92,Percival},
Fokker-Planck or c-number Langevin equations \cite{Walls94,Haken}, quantum
Langevin equations \cite{Walls94,Scully97,Gardiner91} are also used.
Sometimes an apparently non-Markovian problem can be converted to a
Markovian one by a more suitable treatment of the internal system
interactions (for example, the use of dressed atom states \cite
{Cohen-Tannoudji69,Cohen-Tannoudji77} for treating driven atoms in narrow
band squeezed vacuum fields \cite{Yeoman96,DaltonFS99}).

However, situations are now being studied in which the standard Born-Markoff
approach is no longer appropriate, since the reservoir correlation times are
too long for the time scales of interest. A structured rather than flat
reservoir situation applies \cite{Lambropoulos00,Mossberg94}. This includes
cases where the reservoir coupling constants vary significantly with
frequency, such as the interaction of atom(s) or quantum dots with light in
high Q cavities \cite{Kimble94}, including microcavities (see, for example, 
\cite{Wu00,Dung99,An94,Mu92,Yamamoto93}) and microspheres (see, for example, 
\cite{Klimov99,Braginsky89}). Cases such as an atom (or many
atoms---super-radiance) interacting with light in photonic band gap
materials \cite{Lambropoulos00,Woldeyohannes99,Paspalakis99,Vats98},\cite
{Yablonovitch87,John87}, where the reservoir mode densities that have gaps
and non-analytic behaviour near the band gap edges, also occur. Also,
quantum feedback situations \cite{Wiseman93,Wiseman94} can involve
significant time delays in the feedback circuit, and thus result in
non-Markovian dynamics for the system itself. Furthermore, systems with
several degrees of freedom, such as in quantum measurements (for example,
the Stern-Gerlach experiment) could involve situations where the decoherence
times associated with some degrees of freedom (such as the position of the
atomic spin) could become so short that the Markoff condition might no
longer be valid, and the effects of such non-Markovian relaxation on the
decoherence times associated with more important degrees of freedom (atom
spin states) would be of interest.

A number of methods for treating non-Markovian processes have been
developed. Apart from direct numerical simulations \cite{Nikolopoulos},
these include the Zwanzig-Nakajima non-Markovian master equation and its
extensions \cite{Zwanzig,Nakajima,Dalton82}, the time-convolutionless
projection operator master equation \cite{Shibata}, Heisenberg equations of
motion \cite{Vats98,Cresser00}, stochastic wave function methods for
non-Markovian processes \cite
{Breuer99,Strunz99,Jack99,Molmer99,Quang97,Stenius96}, methods based on the
essential states approximation or resolvent operators \cite
{Lambropoulos00,Paspalakis99,Bay97}, the pseudo mode approach \cite
{Garraway97,Imamoglu94}, Fano diagonalisation \cite{Jeffers,Barnett97,Fano61}
and the sudden decoherence approximation \cite{Braun}. Of these methods the
last four are easier to apply and give more physical insight into what is
happening. However the essential states method is difficult to apply in all
but the simplest situations, since the set of coupled amplitude equations
becomes unwieldy and it is difficult to solve---as in the case where
multiple excitations of the reservoir are involved. The pseudo mode approach
is based on the idea of enlarging the system to include part of the
reservoir (the pseudo mode---which could be bosonic or fermionic depending
on the case) thereby forming a bigger system in which the Markoff
approximation now applies when the coupling to the remainder of the
reservoir is treated. At present the pseudomode method is also restricted to
single reservoir excitation cases. The Fano diagonalisation method relates
the causes of non-Markovian effects to various underlying features (such as
the presence of bound states for treating atom lasers), and is closely
related to the pseudo mode method. The sudden decoherence method enables
decoherence effects on time scales short compared to system Bohr periods to
be treated simply via ignoring the system Hamiltonian.

This paper deals with the relationship between the current pseudo mode
method for single quantum reservoir excitations and the Fano diagonalisation
method for situations where the reservoir structure is due to the presence
of a discrete, system of (quasi) modes which are coupled to other continuum
(quasi) modes. This important case applies to atomic systems coupled to the
quantum EM\ field in high Q resonant cavities, such as microspheres or
microcavities. It is shown that the pseudo mode method for single quantum
excitations of the structured reservoir can be obtained by applying the Fano
diagonalisation method to a system featuring a set of discrete quasi modes 
\cite{Dalton99a,Barnett88} together with a set of continuum quasi modes,
whose mode density is slowly varying. The structured reservoir of true modes 
\cite{Dalton96,Lang73} is thus replaced by the quasi modes. The interaction
between the discrete and continuum quasi modes is treated in the
rotating-wave approximation and assuming slowly varying coupling constants 
\cite{Dalton99b,Gardiner85}. The atomic system is assumed to be only coupled
to the discrete quasi modes. The density of continuum quasi modes is
explicitly included in the model. Although the behavior of the atomic system
itself is non-Markovian, the enlarged system obtained by combining the
discrete quasi modes with the atomic system now exhibits Markovian dynamics.
The discrete quasi modes are identified as pseudo modes. The continuum quasi
modes are identified as the flat reservoir to which the enlarged Markovian
system is coupled. Explicit expressions for the atom-true modes coupling
constants are obtained, exhibiting the rapidly varying frequency dependence
characteristic of structured reservoirs. At present the treatment is
restricted to cases where threshold and band gap effects are unimportant,
but may be applicable to two-dimensional photonic band gap materials.
However, the problem of treating multiple excitation processes for certain
types of structured reservoirs can now be treated via the quasimode theory,
since the Markovian master equation for the atom-quasi mode system applies
for cases involving multilevel atoms or cases of several excited two-level
atoms. Further extensions of the treatment to allow for atomic systems
driven by single mode external laser fields are also possible, with the
original atomic system being replaced by the dressed atom.

The plan of the paper is as follows. In Section \ref{SPseudo} the key
features of pseudo mode theory are outlined. Section \ref{SFano} presents
the Fano diagonalisation theory for the quasi mode system, with details
covered in the Appendices \ref{App. Atomic Ham}, \ref{App.Scaling} and \ref
{h=adaga}. In Section \ref{SResults} specific cases such as one or two
discrete quasi modes or where the variation of discrete-continuum coupling
constants can be ignored are examined, giving results for the atom-true mode
coupling constants and reservoir structure functions in these situations.
Section \ref{SMarkov} contains the Markovian master equation for the atom
plus discrete quasi modes system. Section \ref{SVarying}\ briefly examines
the situation where coupling constants and mode densities are not slowly
varying. Conclusions and comments are set out in Section \ref{SConc}.

\section{ Pseudomode Theory}

\label{SPseudo}

\begin{figure}[tbp]
 \centerline{\includegraphics[scale=0.5]{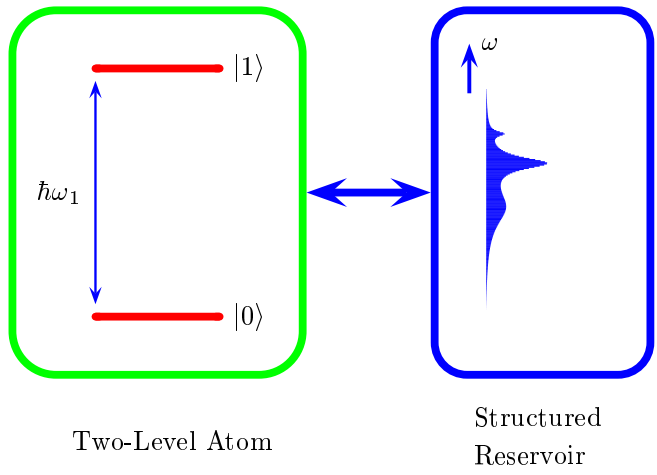}}
 \caption[f1]{ Illustration of a two-level atom coupled to a structured
reservoir. }
\label{Figure 1}
\end{figure}

The simplest case to which pseudomode theory \cite{Garraway97} can be
applied is that of a two-level atom coupled to the modes of the quantum
electromagnetic (EM) field---which constitutes the structured reservoir.
Only one photon excitation processes will occur. However, the formalism
would also apply to any spin 1/2 fermion system coupled to a bath of bosonic
oscillators. The Hamiltonian is given in the rotating wave approximation by: 
\begin{eqnarray}
\hat{H} &=&\sum_{\lambda }\hbar \omega _{\lambda }\hat{a}_{\lambda
}^{\dagger }\hat{a}_{\lambda }+{\small \frac{1}{2}}\hbar \omega _{1}\left( 
\hat{\sigma}^{+}\hat{\sigma}^{-}-\hat{\sigma}^{-}\hat{\sigma}^{+}\right) 
\nonumber \\
&&+\sum_{\lambda }\left( \hbar g_{\lambda }^{\ast }\hat{a}_{\lambda }\hat{%
\sigma}^{+}+\mbox{h.c.}\right)  \label{pseudoham}
\end{eqnarray}
where $\hat{\sigma}^{+},\hat{\sigma}^{-}$ are the usual atomic spin
operators, $\hat{a}_{\lambda }$, $\hat{a}_{\lambda }^{\dagger }$ are the
annihilation, creation operators for the mode $\lambda $ of the field, $%
\omega _{1}$ is the atomic transition frequency, $\omega _{\lambda }$ is the
mode frequency and $g_{\lambda }$ are the coupling constants. This system is
illustrated in Fig. \ref{Figure 1}.

To describe a one photon excitation process, the initial condition is the
atom excited and no photons present in the field. Hence the initial
Schrodinger picture state vector is: 
\begin{equation}
|\Psi (0)\rangle =|1\rangle |...0_{\lambda }...\rangle
\end{equation}

In the essential states approach the state vector at later time $t$ will be
a superposition of the initial state and states with the atom in its lower
state and one photon in various modes $\lambda $. With $\tilde{c}_{1},\tilde{%
c}_{\lambda }$ defining the complex amplitudes for the atomic excited state
and the one photon states, the state vector is: 
\begin{equation}
|\Psi (t)\rangle =\tilde{c}_{1}(t)\mbox{e}^{-i\omega _{1}t}|1\rangle
|...0_{\lambda }...\rangle +\sum_{\lambda }\tilde{c}_{\lambda }(t)\mbox{e}%
^{-i\omega _{\lambda }t}|0\rangle |...1_{\lambda }...\rangle
\label{essentstate}
\end{equation}

Substitution into the time dependent Schrodinger equation leads to the
following coupled complex amplitude equations: 
\begin{eqnarray}
i\frac{\mbox{d}}{\mbox{d}t}\tilde{c}_{1} &=&\sum_{\lambda }g_{\lambda
}^{\ast }\mbox{e}^{-i\Delta _{\lambda }t}\tilde{c}_{\lambda }  \nonumber \\
i\frac{\mbox{d}}{\mbox{d}t}\tilde{c}_{\lambda } &=&g_{\lambda }\mbox{e}%
^{i\Delta _{\lambda }t}\tilde{c}_{1}  \label{coupampeqns}
\end{eqnarray}
where $\Delta _{\lambda }=\omega _{\lambda }-\omega _{1}$ are detunings.

Formally eliminating the amplitudes for the one photon states enables an
integro-differential equation for the excited atomic amplitude to be
derived. This is: 
\begin{equation}
\frac{\mbox{d}}{\mbox{d}t}\tilde{c}_{1}(t)=-\int_{0}^{t}\mbox{d}\tau \tilde{G%
}(\tau )\tilde{c}_{1}(t-\tau )  \label{integrode}
\end{equation}
and involves a kernel $\tilde{G}$ given by 
\begin{eqnarray}
\tilde{G}(\tau ) &=&\sum_{\lambda }|g_{\lambda }|^{2}\mbox{e}^{-i\Delta
_{\lambda }\tau }  \nonumber \\
&=&\int d\omega _{\lambda }\rho (\omega _{\lambda })|g_{\lambda }|^{2}%
\mbox{e}^{-i\Delta _{\lambda }\tau }
\end{eqnarray}
The mode density $\rho (\omega _{\lambda })$ is introduced after replacing
the sum over $\lambda $ by an integral over the mode angular frequency $%
\omega _{\lambda }$.

It is apparent from Eq.(\ref{integrode}) that the behaviour of the atomic
system only depends on the reservoir structure function $D(\omega _{\lambda
})$ for this single quantum excitation case defined by 
\begin{equation}
\rho _{\lambda }\left| g_{\lambda }\right| ^{2}=\frac{\Omega ^{2}}{2\pi }%
D(\omega _{\lambda })
\end{equation}
where a transition strength $\Omega $ is introduced to normalise $D$ so that
its integral gives $2\pi $. The transition strength $\Omega $ is given by: 
\begin{equation}
\Omega ^{2}=\int d\omega _{\lambda }\rho (\omega _{\lambda })|g_{\lambda
}|^{2}
\end{equation}
The reservoir structure function $D(\omega _{\lambda })$ enables us to
describe the various types of reservoir to which the atomic system is
coupled. If $D$ is slowly varying as a function of $\omega _{\lambda }$ then
the reservoir is `flat', whilst `structured' reservoirs are where $D$ varies
more rapidly, as seen in Figure \ref{Figure 1}. There are of course two
factors involved in determining the behaviour of $D$---the mode density $%
\rho (\omega _{\lambda })$ and the coupling constant via $|g_{\lambda }|^{2}$%
. Either or both can determine how structured the reservoir is. Photonic
band gap materials are characterised by mode densities which are actually
zero over the gaps in the allowed mode frequencies, and which have non
analytic behaviours near the edges of the band gaps. All mode densities are
zero for negative $\omega _{\lambda }$, so threshold effects are possible.
In cavity QED situations, such as for microspheres and other high Q
cavities, the coupling constant varies significantly near the cavity
resonant frequencies, so in these cases it is the coupling constant that
gives structure to the reservoir. For the present it will be assumed that
(apart from simple poles) $D$ is analytic in the lower half complex $\omega
_{\lambda }$ plane, and any other non analytic features can be disregarded.
It is recognized of course that this restriction places a limit on the range
of applicability of the theory, though it may be possible to extend this
range by representing the actual function $D(\omega _{\lambda })$ in an
approximate form that satisfies the analyticity requirements. In addition
(in order to calculate the contour integrals) it will be assumed that $D$
tends to zero at least as fast as $1/|\omega _{\lambda }|$ as $|\omega
_{\lambda }|$ tends to infinity.

Based on the above assumption regarding the reservoir structure function $D$%
, the kernel $G$ may be evaluated in terms of the poles and residues of $%
D(\omega _{\lambda })$ in the lower half complex $\omega _{\lambda }$ plane.
It is assumed that these simple poles can be enumerated. The poles are
located at $z_{1},z_{2},....z_{l}...$ and their residues are $%
r_{1},r_{2},...r_{l}...$. The pole $z_{l}$ may be expressed in terms of a
real angular frequency $\omega _{l}$ and a width factor $\Gamma _{l}$ via $%
z_{l}=\omega _{l}-i\Gamma _{l}/2$. Contour integration methods show that the
sum of the residues equals $i$. The kernel is obtained in the form: 
\begin{equation}
\tilde{G}(\tau )=-i\Omega ^{2}\sum_{l}r_{l}e^{-i(z_{l}-\omega _{1})\tau }
\label{kernelG}
\end{equation}

The integro-differential equation (\ref{integrode}) for the excited atomic
amplitude involves a convolution integral on the right hand side and may be
solved using Laplace transform methods. The atomic behaviour obtained is
well known \cite{Dalton96,Lang73} and will not be rederived here. It is
found that there are two regimes, depending on the ratio of the transition
strength to typical width factors. These are: (a) a strong coupling regime
with Non-Markovian atomic dynamics, and occuring when $\Omega \gg \Gamma $
(b) a weak coupling regime with Markovian atomic dynamics, occuring when $%
\Omega \ll \Gamma $.

The pseudomode approach continues by considering poles of the reservoir
structure function $D(\omega _{\lambda })$ in the lower half complex $\omega
_{\lambda }$ plane. Each pole will be associated with one pseudomode.
Reverting to Schrodinger picture amplitudes via $c_{1}(t)=\tilde{c}_{1}(t)%
\mbox{e}^{-i\omega _{1}t}$ etc., pseudomode amplitudes associated with each
pole of $D$ are introduced as defined by: 
\begin{equation}
b_{l}(t)=-i\Omega \sqrt{-ir_{l}}e^{-iz_{l}t}\int_{0}^{t}\mbox{d}t^{\prime
}e^{iz_{l}t^{\prime }}c_{1}(t^{\prime })  \label{pseudoamp}
\end{equation}

 From the definition of $b_{l}$ and by substituting the form (Eq.(\ref
{kernelG})) for the kernel $G$ that involves the poles of $D$, it is not
difficult to show that the excited atomic amplitude and the pseudo mode
amplitudes satisfy the following coupled equations: 
\begin{eqnarray}
i\frac{dc_{1}(t)}{dt} &=&\omega _{1}c_{1}(t)+\sum_{l}{\cal K}_{l}b_{l}(t) 
\nonumber \\
i\frac{db_{l}(t)}{dt} &=&z_{l}b_{l}(t)+{\cal K}_{l}c_{1}(t)
\label{pseudoeqns}
\end{eqnarray}
where ${\cal K}_{l}=\Omega \sqrt{-ir_{l}}$ are pseudomode coupling
constants. In general, the residues $r_{l}$ are not pure imaginary, so the
pseudomode coupling constants are not real.

The important point is that the atom plus pseudomodes system now satisfies
Markovian equations (Eq.(\ref{pseudoeqns})). With a finite (or countable)
set of pseudomodes, the original atom plus structured continuum has now been
replaced by a simpler system which still enables an exact description of the
atomic behaviour to be obtained. {\em Exact} master equations involving the
pseudomodes have been derived, and in general the pseudomodes can be coupled
(see \cite{Garraway97}). There are however, difficulties in cases where the
pseudomode coupling constants are not real, which can occur in certain cases
where there are several pseudomodes. Apart from the general difficulty
associated with situations where (apart from having simple poles) the
reservoir structure function $D$ is not analytic in the lower half complex
plane, there is also considerable difficulty in extending the above theory
to treat cases where multiple excitations of the structured reservoir occur,
such as when the two-level atom is replaced by a three-level atom in a
cascade configuration with the atom is initially in the topmost state. The
problem is that applying the usual essential states approach leads to two
(or more) photon states now appearing in the state vector (cf. Eq.(\ref
{essentstate})), and the resulting coupled amplitude equations (cf. Eq.(\ref
{coupampeqns})) do not apear to facilitate the sucessive formal elimination
of the one, two, .. photon amplitudes, as is possible in the single photon
excitation case treated above. It is therefore not clear how pseudomode
amplitudes can be introduced, along the lines of Eq.(\ref{pseudoamp}) so the
pseudomode method has not yet been generalised from its original formulation
to allow for multiple reservoir excitations.

%%%%%%%%%%%%%%%%%%%%%%%%%%%%%%%%%%%%%%%%%%%%%%%%%%%%%%%%%%%%%%5

\section{ Fano Diagonalisation for a Quasi Mode System}

\label{SFano}

\subsection{ Description of the approach}

\label{SubSDescription}

The case of multiple excitation of a structured reservoir involves systems
more complex than the two level atom treated above. It will be sufficient
for the purpose of linking the pseudomode and Fano diagonalisation methods
to consider single multilevel atomic systems, although multiatom systems
would also be suitable as both systems could result in multiphoton
excitations of the quantum EM field. Accordingly the two level Hamiltonian
given as the second term in Eq.(\ref{pseudoham}) is now replaced by the
multilevel atomic Hamiltonian:

\begin{equation}
\hat{H}_{A}=\sum_{k}\eta _{k}\hbar \omega _{k}\left( \hat{\sigma}_{k}^{+}%
\hat{\sigma}_{k}^{-}-\hat{\sigma}_{k}^{-}\hat{\sigma}_{k}^{+}\right)
\label{atomham}
\end{equation}
The index $k$ represents an atomic transition associated with a pair of
energy levels ($k\equiv \{u,l\}$) with energy difference $\hbar \omega _{k}$%
. The quantities $\eta _{k}$ are numbers chosen so that $\hat{H}_{A}$ equals
the atomic Hamiltonian, apart from an additive constant energy. For example,
in a two level atom $\eta $ $=\frac{1}{2}$ for the single transition, whilst
in a three level atom in a V configuration with degenerate upper levels $%
\eta _{1}$ $=\eta _{2}=\frac{1}{3}$ for the two optical frequency
transitions, and $\eta _{3}=0$ for the zero frequency transition. Details
are set out in Appendix \ref{App. Atomic Ham}. The atomic transition
operators are $\hat{\sigma}_{k}^{+}\equiv |u\rangle \langle l|\equiv (\hat{%
\sigma}_{k}^{-})^{\dagger }$. As the Hamiltonians for other fermionic
systems can also be written in the same form as in Eq.(\ref{atomham}), the
treatment is not just restricted to single multilevel atom systems.

\begin{figure}[tbp]
  \centerline{\includegraphics[scale=0.5]{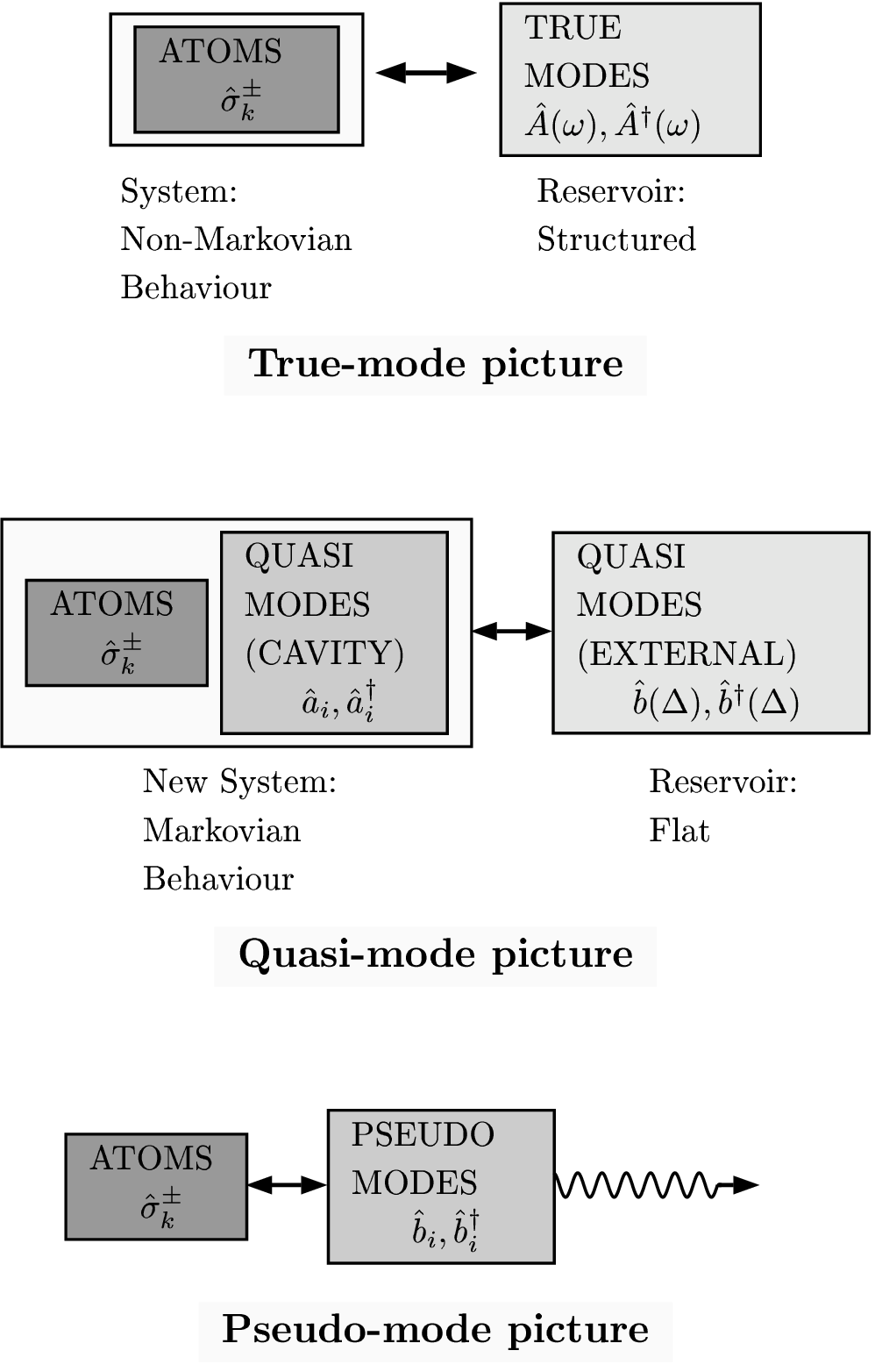}}
\caption[f1]{ Three pictures of the coupled atomic system. In the {\em %
true-mode picture} the atom is coupled directly to true modes which have
structure. In the {\em quasi-mode picture} the atoms are coupled to
quasi-modes which are in turn coupled to external quasi-modes. In the {\em %
pseudo-mode picture} the atoms are coupled to dissipative pseudomodes. }
\label{Figure 2}
\end{figure}

As indicated in Section \ref{SPseudo}, an important pseudo mode situation is
where the reservoir structure is due to the presence of a discrete, system
of (quasi) modes which are coupled to other continuum (quasi) modes with
slowly varying coupling constants. This important case applies to atomic
systems coupled to the quantum EM\ field in high Q resonant cavities, such
as microspheres or microcavities. The Fano diagonalisation method is then
based around the idea that the structured reservoir of quantum EM\ field
modes can be described in two different ways, which will now be outlined.\
Figure \ref{Figure 2} illustrates these two descriptions, along with that
involving pseudo modes.

\subsubsection{Quasi modes}

\label{SubSubQuasi}

The first approach is to treat the quantum EM\ field in terms of a quasi
mode description \cite{Dalton99a,Barnett88}. The quasi modes behave as
coupled quantum harmonic oscillators. These are to consist of two types, the
first being a set of discrete quasi modes, the second being a set of
continuum quasi modes. In a typical structured reservoir situation for the
area of cavity QED \cite{Dalton99b}, the quasi modes represent a realistic
description of the physical system. The discrete modes would be cavity quasi
modes---one for each cavity resonance and appropriate for describing the EM\
field inside the cavity, the continuum modes would be external quasi modes
which are describe the field outside the cavity. The interaction between the
discrete and continuum quasi modes will be treated in the rotating-wave
approximation assuming slowly varying coupling constants \cite
{Dalton99b,Barnett88,Gardiner85}. Rotating-wave approximation couplings
between the discrete quasi modes are also included, but couplings between
the continuum quasi modes are not included---such couplings can be removed
by pre-diagonalisation. For the quasi mode description the field Hamiltonian
is given by:

\begin{eqnarray}
\hat{H}_{F} &=&\sum_{i}\hbar \omega _{i}\hat{a}_{i}^{\dagger }\hat{a}%
_{i}+\sum_{i\neq j}\hbar v_{ij}\hat{a}_{i}^{\dagger }\hat{a}_{j}  \nonumber
\\
&&+\sum_{i}\int d\Delta \rho _{c}(\Delta )\left[ \hbar W_{i}(\Delta )\hat{a}%
_{i}^{\dagger }\hat{b}(\Delta )+\mbox{h.c.}\right]  \nonumber \\
&&+\int d\Delta \rho _{c}(\Delta )\,\hbar \Delta \hat{b}^{\dagger }(\Delta )%
\hat{b}(\Delta )  \label{quasifieldham}
\end{eqnarray}
where $\hat{a}_{i}$, $\hat{a}_{i}^{\dagger }$ are the annihilation, creation
operators for the discrete quasi mode $i$, $\omega _{i}$ is its frequency, $%
\hat{b}(\Delta ),\hat{b}^{\dagger }(\Delta )$ are the annihilation, creation
operators for the continuum quasi mode of frequency $\Delta $, the coupling
constants between the $i,j$ discrete quasi modes are $v_{ij}$ ($%
v_{ij}=v_{ji}^{\ast }$), whilst the quantity $W_{i}(\Delta )$ is the
coupling constant between the $i$ discrete and $\Delta $ continuum quasi
modes. The integrals over the quasi continuum frequency $\Delta $ involve a
quasi continuum mode density $\rho _{c}(\Delta )$. Both $W_{i}(\Delta )$ and 
$\rho _{c}(\Delta )$ are usually slowly varying. The discrete quasi mode
annihilation, creation operators satisfy Kronecker delta commutation rules,
whilst those for the continuum quasi mode operators satisfy Dirac delta
function commutation rules:

\begin{eqnarray}
\lbrack \hat{a}_{i},a_{j}^{\dagger }] &=&\delta _{ij}  \nonumber \\
\lbrack \hat{b}(\Delta ),\hat{b}^{\dagger }(\Delta ^{\prime })] &=&\delta
(\Delta -\Delta ^{\prime })/\rho _{c}(\Delta ).  \label{quasicomm}
\end{eqnarray}
The $\rho _{c\text{ }}$factor on the right hand side gives annihilation and
creation operators which are dimensionless.

For the quasi mode description the interaction between the atomic system and
the quantum EM\ field will be given in the rotating-wave approximation and
only involve coupling to the discrete quasi modes. This would apply for the
typical structured reservoir situation for the area of cavity QED in the
familiar case where the atoms are located inside the cavity. The energy of
an excited atom escapes to the external region in a two step process: first,
a photon is created in a discrete (cavity) quasi mode via the atom-discrete
quasi mode interaction, then second, this photon is destroyed and a photon
is created in a continuum (external) quasi mode via the discrete-continuum
quasi mode coupling. For the quasi mode description the atom-field
interaction will be given as:

\begin{equation}
\hat{H}_{AF}=\sum_{k}\,\sum_{i}\,\left( \hbar \lambda _{ki}^{\ast }\hat{a}%
_{i}\hat{\sigma}_{k}^{+}+\mbox{h.c.}\right)  \label{quasiatomfieldintn}
\end{equation}
where $\lambda _{ki}$ is the coupling constant for the $k$ atomic transition
and the $i$ quasi mode.

\subsubsection{True modes}

\label{SubSubTrue}

The second way of describing the quantum EM\ field is in terms of its true
modes \cite{Dalton96,Lang73}. The true modes behave as uncoupled quantum
harmonic oscillators. These modes are also used in cavity QED and are often
referred to as ``universe modes''. The pseudomode theory presented in
Section \ref{SPseudo} is also based on true modes. For frequencies near to
the cavity resonances these modes are large inside the cavity and small
outside, for frequencies away from resonance the opposite applies. The
distinction between true modes and quasi modes is discussed in some detail
in recent papers \cite{Dalton99a,Brown00} and their detailed forms and
features in the specific case of a planar Fabry-Perot cavity are
demonstrated in \cite{Dalton99b} In terms of true modes the field
Hamiltonian is now given in the alternative form as:

\begin{equation}
\hat{H}_{F}=\int d\omega \,\rho (\omega )\hbar \omega \hat{A}^{\dagger
}(\omega )\hat{A}(\omega )  \label{truefieldham}
\end{equation}
where $\hat{A}(\omega ),$\ $\hat{A}^{\dagger }(\omega )$\ are the
annihilation, creation operators for the continuum true mode of frequency $%
\omega $. The integrals over the quasi continuum frequency $\omega $ involve
the true continuum mode density $\rho (\omega )$, which is not in general
the same function as $\rho _{c}(\Delta )$. It is also not necessarily a
slowly varying function of $\omega $. The continuum true mode annihilation,
creation operators satisfy Dirac delta function commutation rules: 
\begin{equation}
\lbrack \hat{A}(\omega ),\hat{A}^{\dagger }(\omega ^{\prime })]=\delta
(\omega -\omega ^{\prime })/\rho (\omega )  \label{truecomm}
\end{equation}
In all these Hamiltonians the coupling constants have dimensions of
frequency, whilst the annihilation and creation operators are dimensionless,
as are the atomic transition operators.

\subsubsection{Relating quasi and true modes}

\label{SubSubQT}

As will be demonstrated in Section \ref{SubDiagonalisation}, Fano
diagonalisation involves determining the relationship between the true mode
annihilation operators $\hat{A}(\omega )$ and the quasi mode annihilation
operators $\hat{a}_{i}$ and $\hat{b}(\Delta )$. The $\hat{A}(\omega )$ will
be written as a linear combination of the $\hat{a}_{i}$ (sum over $i$) and $%
\hat{b}(\Delta )$ (integral over $\Delta $)(see Eq.(\ref{A.def}) below),
which involves the functions $\alpha _{i}(\omega )$ and $\beta (\omega
,\Delta )$. This relationship can be inverted to give the $\hat{a}_{i}$ as
an integral over $\omega $ of the $\hat{A}(\omega )$ (see Eq.(\ref{ab})
below). This enables the true mode form of the atom-field interaction to be
given as: 
\begin{equation}
\hat{H}_{AF}=\sum_{k}\sum_{i}\int d\omega \,\rho (\omega )\left( \hbar
\lambda _{ki}^{\ast }\alpha _{i}^{\ast }(\omega )\hat{A}(\omega )\hat{\sigma}%
_{k}^{+}+\mbox{h.c.}\right)  \label{trueatomfieldintn}
\end{equation}
Comparing Eqs.(\ref{quasiatomfieldintn}) and (\ref{trueatomfieldintn}) we
see that the atom-true mode coupling constant $g^{k}(\omega )$ (for the $k$
atomic transition and the $\omega $ true mode) is given by the expression: 
\begin{equation}
g^{k}(\omega )=\sum_{i}\lambda _{ki}\alpha _{i}(\omega ).
\label{truemodecouplingconst}
\end{equation}
This can be a complicated function of $\omega $ in a structured reservoir,
as will be seen from the forms obtained for the function $\alpha _{i}(\omega
)$ (for example, Eq.(\ref{alphai})). This expression for the atom-true mode
coupling constant is one of the key results in our theory, and enables the
pseudomode and quasi mode descriptions of decay processes for structured
reservoirs to be related. Note that the true mode coupling constant now
involves two factors: the atom-quasi mode coupling constant $\lambda _{ki}$
and the function $\alpha _{i}(\omega )$ that arises from the Fano
diagonalisation process.

For the situation where only a single atomic transition $k$ is involved, the
equivalent reservoir structure function would be given by:

\begin{equation}
D^{k}(\omega )=C\rho (\omega )|g^{k}(\omega )|^{2}  \label{reservfn2}
\end{equation}
where $C$ is the normalising constant, which for convenience we will set
equal to unity as it does not contain any $\omega $ dependence. This
expression will be used to compare the results from the quasi mode approach
to those of the present single quantum excitation pseudomode theory. As we
will see, the true mode density cancels out.

Finally, athough our results are still correct for cases where the quasi
mode density $\rho _{c}(\Delta )$ and the coupling constants $W_{i}(\Delta )$
are not restricted to being slowly varying functions of $\Delta $, their
utility where this is not the case is somewhat limited. The theory is mainly
intended to apply to the important pseudo mode situation where the reservoir
structure is actually due to the presence of a discrete, system of quasi
modes which are coupled to other continuum quasi modes via slowly varying
coupling constants. For example, the quantum EM\ field in high Q resonant
cavities can be accurately described in terms of the quasi mode model which
has these features, the discrete quasi modes being the cavity quasi modes
(linked to the cavity resonances) with which the atoms inside the cavity
interact, and the continuum quasi modes being the external modes.

As pointed out previously, the structured reservoir can be any set of
bosonic oscillators, not just the quantum EM\ field. The above treatment
would thus apply more generally, and we would then refer to discrete quasi
oscillators, continuum quasi oscillators or true oscillators. The physical
basis for a quasi mode description of the reservoir of bosonic oscillators
will depend on the particular situation; in general they will be idealised
approximate versions of the true modes.

%%%%%%%%%%%%%%%%%%%%%%%%%%%%%%%%%%%%%%%%%%%%%%%%%%%%%%%%%%%

\subsection{ Diagonalisation of the quasi mode Hamiltonian: dressing the
quasi mode operators}

\label{SubDiagonalisation}

\subsubsection{Basic equations for Fano diagonalisation}

We start with a multiple quasi-mode description of the quantum EM\ field,
for which the Hamiltonian is given above as Eq.(\ref{quasifieldham}). This
Hamiltonian can also be written in terms of the true mode description as in
Eq. (\ref{truefieldham}), and the problem is to relate the true mode
annihilation operators $\hat{A}(\omega )$ in terms of the quasi mode
annihilation operators $\hat{a}_{i}$ and $\hat{b}(\Delta )$. In view of the
rotating wave approximation form of the Hamiltonian, the quasi mode creation
operators are not involved in the relationship \cite{Dalton99a}. Fano
diagonalisation for the non-rotating wave approximation has been treated for
the case of a single mode coupled to a reservoir in Ref.\cite
{Huttner92,Rosenau00}. In making a Fano diagonalization we will follow the
lines of Ref.\cite{Barnett97} (Section 6.6 on dressed operators), rather
than Ref.\cite{Fano61}, but note that a new feature here is the presence of
the mode-mode coupling term in the Hamiltonian Eq.(\ref{quasifieldham}). In
addition, we explicitly include the mode densities from the beginning. The
physical realisation of the quasi mode model for the EM\ field really
determines the quasi continuum mode density $\rho _{c}(\Delta )$, just as it
does the coupling constants $v_{ij},W_{i}(\Delta )$ and $\lambda _{ki}$. It
is therefore important to be able to find the $\rho _{c}(\omega )$
dependence of quantities such as the reservoir structure function $D(\omega
) $ (as we will see, the final expression (Eq.(\ref{reservfn6})) for the
latter does not involve the true mode density $\rho (\omega )$). It is of
course possible to scale all the other quantities to make $\rho =\rho _{c}=1$%
, and then rescale afterwards to allow for the actual $\rho ,\rho _{c}$ that
apply for the system of interest, but this would lead to much duplication of
the results we present. For completeness, the scaling is set out in Appendix 
\ref{App.Scaling}.

 From the form of the true mode Hamiltonian in Eq.(\ref{truefieldham})\ and
the commutation rules Eq.(\ref{truecomm}) to be satisfied by the $\hat{A}%
(\omega )$, it is clear that the true mode annihilation operators are
eigenoperators of the quantum field Hamiltonian $\hat{H}_{F}$ and must
satisfy: 
\begin{equation}
\left[ \hat{A}(\omega ),\hat{H}_{F}\right] =\hbar \omega \hat{A}(\omega )
\label{true.def}
\end{equation}

In general, the true mode annihilation operators $\hat{A}(\omega )$ can be
expressed as linear combinations of the quasi mode annihilation operators $%
\hat{a}_{i}$ and $\hat{b}(\Delta )$ in the form (\cite{Barnett88,Dalton99a}%
): 
\begin{equation}
\hat{A}(\omega )=\sum_{i}\alpha _{i}(\omega )\hat{a}_{i}+\int \,d\Delta \rho
_{c}(\Delta )\beta (\omega ,\Delta )\hat{b}(\Delta ),  \label{A.def}
\end{equation}
where $\alpha _{i}(\omega )$ and $\beta (\omega ,\Delta )$ are functions to
be determined, and which are dimensionless. This form for $\hat{A}(\omega )$
is then substituted into Eq.(\ref{true.def}) and the commutator evaluated
using the quasi mode form Eq.(\ref{quasifieldham}) for $\hat{H}_{F}$ and the
commutation rules in Eq.(\ref{quasicomm}). The coefficients of the the
operators $\hat{a}_{i}$ and $\hat{b}(\Delta )$ on both sides of Eq.(\ref
{true.def}) are then equated, giving a set of coupled equations for the $%
\alpha _{i}(\omega )$ and $\beta (\omega ,\Delta )$. These are: 
\begin{eqnarray}
(\omega _{i}-\omega )\alpha _{i}(\omega )+\sum_{j\neq i}v_{ji}\alpha
_{j}(\omega )+\int \,d\Delta \rho _{c}(\Delta )\beta (\omega ,\Delta
)W_{i}^{\ast }(\Delta ) &=&0  \label{alpha.eq} \\
(\Delta -\omega )\beta (\omega ,\Delta )+\sum_{i}W_{i}(\Delta )\alpha
_{i}(\omega ) &=&0  \label{beta.eq}
\end{eqnarray}

To solve Eqs.(\ref{alpha.eq}), (\ref{beta.eq}) for the unknown $\alpha
_{i}(\omega )$ and $\beta (\omega ,\Delta )$ we first solve for $\beta $ in
terms of the $\alpha _{i}$. This gives: 
\begin{equation}
\beta (\omega ,\Delta )=\left[ \frac{{\cal P}}{\omega -\Delta }+z(\omega
)\delta (\omega -\Delta )\right] \sum_{j}W_{j}(\Delta )\alpha _{j}(\omega ),
\label{beta.def}
\end{equation}
where $z(\omega )$ is a dimensionless function yet to be determined. This
expression is then substituted into Eq.(\ref{alpha.eq}) to obtain a set of
linear homogeneous equations for the $\alpha _{i}(\omega )$ in the form: 
\begin{equation}
(\omega _{i}-\omega )\alpha _{i}(\omega )+\sum_{j\neq i}v_{ji}\alpha
_{j}(\omega )+\sum_{j}F_{ij}\alpha _{j}(\omega )+\sum_{j}W_{i}^{\ast
}(\omega )W_{j}(\omega )\rho _{c}(\omega )z(\omega )\alpha _{j}(\omega )=0.
\label{z1}
\end{equation}
In these equations, a frequency shift matrix $F_{ij}(\omega )$ appears,
which involves a principal integral of products of the discrete-continuum
quasi mode coupling constants together with the quasi continuum mode
density. This is defined by: 
\begin{equation}
F_{ij}(\omega )={\cal P}\int d\Delta \rho _{c}(\Delta )\frac{W_{i}^{\ast
}(\Delta )W_{j}(\Delta )}{\omega -\Delta }  \label{Fij.def}
\end{equation}
and satisfies the Hermiticity condition $F_{ji}=F_{ij}^{\ast }.$

Equation (\ref{z1}) can be written in the matrix form 
\begin{equation}
{\bf m}\bbox{ \alpha}=0  \label{m1}
\end{equation}
where the column matrix $\bbox{\alpha}\equiv \{\alpha _{1}(\omega ),\alpha
_{2}(\omega ),\alpha _{3}(\omega ),...\}^{\top }$ and the square matrix $%
{\bf m}$ is given by: 
\begin{equation}
{\bf m}_{ij}(\omega )=(\omega _{i}-\omega )\delta _{ij}+(1-\delta
_{ij})v_{ji}+F_{ij}(\omega )+W_{i}^{\ast }(\omega )W_{j}(\omega )\rho
_{c}(\omega )z(\omega ).  \label{m.def}
\end{equation}

\subsubsection{Solution of equations for amplitudes $\protect\alpha _{i}(%
\protect\omega )$ and $\protect\beta (\protect\omega ,\Delta )$}

The approach used to solve these equations is as follows. It is clear that
Eq.(\ref{m1}) can give an (unnormalized) solution for $\bbox{\alpha}$ in
terms of the function $\rho _{c}(\omega )z(\omega )$. We can now use Eq.(\ref
{m1}) itself to obtain the expression for $\rho _{c}(\omega )z(\omega )$,
subject to the assumption that the quantity $\sum_{i}W_{i}(\omega )\alpha
_{i}(\omega )$ is non-zero. This assumption will be verified {\it a
posteriori} from the normalisation condition for the $\alpha _{i}(\omega )$,
which will follow (see below) from the requirement that the form for the $%
\hat{A}(\omega )$ given in Eq.(\ref{A.def}), satisfies the commutator
relation $[\hat{A}(\omega ),\hat{A}^{\dagger }(\omega ^{\prime })]=\delta
(\omega -\omega ^{\prime })/\rho (\omega )$, [Eq.(\ref{truecomm})]. This
indeed leads to a non-zero expression for $\sum_{i}W_{i}(\omega )\alpha
_{i}(\omega )$, (see Eq.(\ref{anorm3}) below). After finding both $\rho
_{c}(\omega )z(\omega )$ and $\sum_{i}W_{i}(\omega )\alpha _{i}(\omega )$
the results can be substituted back into the equations (\ref{z1}). By
eliminating the factor $\sum_{i}W_{i}(\omega )\alpha _{i}(\omega )$ from the
last term in Eqs.(\ref{z1}), we obtain a set of inhomogeneous linear
equations for the $\alpha _{i}(\omega )$, which can then be solved for the $%
\alpha _{i}(\omega )$ (and hence $\beta (\omega ,\Delta )$).

The general expression for $\rho _{c}(\omega )$ $z(\omega )$ can be obtained
from the matrix equation (\ref{m1}). With ${\bf E}$ the unit matrix we
introduce the square matrix ${\bf \Omega }$, the column matrix ${\bf W}^{%
{\bf \ast }}$ and the row matrix ${\bf W}^{T}$ via:

\begin{equation}
{\bf \Omega }_{ij}(\omega )=\omega _{i}\delta _{ij}+(1-\delta
_{ij})v_{ji}+F_{ij}(\omega )  \label{omega.def}
\end{equation}
and ${\bf W}^{\ast }(\omega )\equiv $ $\{W_{1}^{\ast }(\omega ),W_{2}^{\ast
}(\omega ),W_{3}^{\ast }(\omega ),...\}^{\top }$, ${\bf W}^{T}(\omega
)\equiv $ $\{W_{1}(\omega ),W_{2}(\omega ),W_{3}(\omega ),...\}$, and then
write Eq.(\ref{m1}) in the form:

\begin{equation}
(-(\omega {\bf E}-{\bf \Omega })+\rho _{c}(\omega )z(\omega ){\bf W}^{\ast }%
{\bf W}^{T})\bbox{\alpha}=0.  \label{m2}
\end{equation}
Now the matrix ${\bf \Omega }$ is Hermitean and positive definite, having
real eigenvalues close to the real and positive $\omega _{i}$. The matrix $%
\omega {\bf E}-{\bf \Omega }$ can be hence assumed to be invertible, so by
multiplying Eq.(\ref{m2}) from the left by ${\bf W}^{T}(\omega {\bf E}-{\bf %
\Omega })^{-1}$ we see that:

\begin{equation}
(-1+\rho _{c}(\omega )z(\omega )J(\omega )){\bf W}^{T}\bbox{\alpha}=0
\end{equation}
where the function $J(\omega )$ is defined by

\begin{equation}
J(\omega )={\bf W}^{T}(\omega {\bf E}-{\bf \Omega })^{-1}{\bf W}^{\ast }.
\label{J.def}
\end{equation}
Now the quantity ${\bf W}^{T}\bbox{\alpha}$ is equal to $\sum_{i}W_{i}(%
\omega )\alpha _{i}(\omega )$, which is assumed to be non zero for reasons
explained above. This means that $(-1+\rho _{c}(\omega )z(\omega )J(\omega
))=0$, and this gives for $\rho _{c}(\omega )z(\omega )$ the general result:

\begin{equation}
\rho _{c}(\omega )z(\omega )=\left\{\sum_{ij}W_{i}(\omega )(\omega {\bf E}-%
{\bf \Omega (}\omega {\bf )})_{ij}^{-1}W_{j}^{\ast }(\omega )\right\}^{-1},
\label{zresult.eq}
\end{equation}
which only involves the various coupling constants and angular frequencies,
along with the quasi continuum mode density. In general the $\omega $
dependence of the result for $\rho _{c}(\omega )z(\omega )$ is complicated,
since both the coupling constants $W_{i}$ and the matrix ${\bf \Omega }$
(via the matrix $F$) will depend on $\omega $. In some important cases
however, their $\omega $ dependence can be ignored.

As indicated previously, Eqs.(\ref{z1}) or (\ref{m1}) only determine the $%
\alpha _{i}(\omega )$ (and hence $\beta (\omega ,\Delta )$) to within an
arbitary scaling factor, as can be seen from their linear form. The
normalisation of the solutions is fixed by noting that we need $\hat{A}%
(\omega )$, Eq.(\ref{A.def}), to satisfy the commutator relation $[\hat{A}%
(\omega ),\hat{A}^{\dagger }(\omega ^{\prime })]=\delta (\omega -\omega
^{\prime })/\rho (\omega )$, [Eq.(\ref{truecomm})]. This leads to the
condition: 
\begin{equation}
\sum \alpha _{i}(\omega )\alpha _{i}^{\ast }(\omega ^{\prime })+\int d\Delta
\rho _{c}(\Delta )\beta (\omega ,\Delta )\beta ^{\ast }(\omega ^{\prime
},\Delta )=\delta (\omega -\omega ^{\prime })/\rho (\omega ).  \label{anorm1}
\end{equation}
Then substituting for $\beta (\omega ,\Delta )$ from Eq.(\ref{beta.def}) and
using Eq.(\ref{Fij.def}), we find after considerable algebra that 
\begin{eqnarray}
\sum \alpha _{i}(\omega )\alpha _{i}^{\ast }(\omega ^{\prime })+\delta
(\omega -\omega ^{\prime })\left( \pi ^{2}+|z(\omega )|^{2}\right) \rho
_{c}(\omega )\sum_{ij}W_{i}(\omega )W_{j}^{\ast }(\omega )\alpha _{i}(\omega
)\alpha _{j}^{\ast }(\omega ) &&  \nonumber \\
+\frac{{\cal P}}{\omega -\omega ^{\prime }}\sum_{ij}\alpha _{i}(\omega
)\alpha _{j}^{\ast }(\omega ^{\prime })\left[ F_{ij}^{\ast }(\omega ^{\prime
})-F_{ji}(\omega )\right. &&  \nonumber \\
\left. +z^{\ast }(\omega ^{\prime })\rho _{c}(\omega ^{/})W_{i}(\omega
^{\prime })W_{j}^{\ast }(\omega ^{\prime })-z(\omega )\rho _{c}(\omega
)W_{i}(\omega )W_{j}^{\ast }(\omega )\right] &=&\delta (\omega -\omega
^{\prime })/\rho (\omega ).  \label{anorm2}
\end{eqnarray}
Note that we have used certain properties of the principal parts and delta
functions (see, for example\ Ref.\ \cite{Barnett97}): 
\begin{eqnarray}
\delta (\omega -\Delta )\delta (\omega ^{\prime }-\Delta ) &=&\delta (\omega
-\omega ^{\prime })\delta (\omega -\Delta )=\delta (\omega -\omega ^{\prime
})\delta (\omega ^{\prime }-\Delta )  \nonumber \\
\frac{{\cal P}}{\omega ^{\prime }-\Delta }\delta (\omega -\Delta ) &=&\frac{%
{\cal P}}{\omega ^{\prime }-\omega }\delta (\omega -\Delta )  \nonumber \\
\frac{{\cal P}}{\omega -\Delta }\,\cdot \,\frac{{\cal P}}{\omega ^{\prime
}-\Delta } &=&\frac{{\cal P}}{\omega -\omega ^{\prime }}\left( \frac{{\cal P}%
}{\omega ^{\prime }-\Delta }-\frac{{\cal P}}{\omega -\Delta }\right) +\pi
^{2}\delta (\omega -\Delta )\delta (\omega ^{\prime }-\Delta )
\label{delta_tech1}
\end{eqnarray}
to obtain the last equation. We then also use 
\begin{equation}
\frac{{\cal P}}{\omega -\omega ^{\prime }}\,\cdot \,(\omega -\omega ^{\prime
})=1  \label{delta_tech2}
\end{equation}
along with Eq.(\ref{z1}) to substitute for $\sum_{j}F_{ij}^{\ast }(\omega
^{\prime })\alpha _{j}^{\ast }(\omega ^{\prime })$ and $\sum_{i}F_{ji}(%
\omega )\alpha _{i}(\omega )$ and obtain finally: 
\begin{equation}
\sum_{ij}W_{i}(\omega )W_{j}^{\ast }(\omega )\alpha _{i}(\omega )\alpha
_{j}^{\ast }(\omega )=\left| \sum_{i}W_{i}(\omega )\alpha _{i}(\omega
)\right| ^{2}=\frac{1}{\rho (\omega )\rho _{c}(\omega )(\pi ^{2}+|z(\omega
)|^{2})}.  \label{anorm3}
\end{equation}
This fixes, albeit with the coefficients $W_{i}(\omega )$, the normalization
of the $\alpha _{i}(\omega )$. Note the appearance of both mode densities in
the result. Finally, with a suitable choice of the overall phase we can fix
the result for the important quantity $\sum_{i}W_{i}(\omega )\alpha
_{i}(\omega )$ to be:

\begin{equation}
\sum_{i}W_{i}(\omega )\alpha _{i}(\omega )=\frac{1}{\sqrt{\rho (\omega )\rho
_{c}(\omega )}(\pi +iz(\omega ))}  \label{anorm6}
\end{equation}

Having obtained this result for $\sum_{i}W_{i}(\omega )\alpha _{i}(\omega )$
we then substitute back into the equations (\ref{z1}), eliminating this
factor from the last term to give a set of inhomogeneous linear equations
for the $\alpha _{i}(\omega )$:

\begin{equation}
(\omega -\omega _{i})\alpha _{i}(\omega )-\sum_{j\neq i}v_{ji}\alpha
_{j}(\omega )-\sum_{j}F_{ij}\alpha _{j}(\omega )=\frac{W_{i}^{\ast }(\omega
)\rho _{c}(\omega )z(\omega )}{\sqrt{\rho (\omega )\rho _{c}(\omega )}(\pi
+iz(\omega ))}.
\end{equation}
After some algebra, introducing the matrix ${\bf \Omega }(\omega )$ from Eq.(%
\ref{omega.def}) and then substituting from Eq.(\ref{zresult.eq})\ for $%
(\rho _{c}(\omega )z(\omega ))^{-1}$, the last equations can be solved for
the $\alpha _{i}(\omega )$, giving the solution in matrix form as:

\begin{equation}
\bbox{\alpha}(\omega ){\bf =-}i\sqrt{\frac{\rho _{c}(\omega )}{\rho (\omega )%
}}\frac{1}{(1-i\pi \rho _{c}(\omega ){\bf W}^{T}(\omega )(\omega {\bf E}-%
{\bf \Omega }(\omega ))^{-1}{\bf W}^{\ast }(\omega ))}(\omega {\bf E}-{\bf %
\Omega }(\omega ))^{-1}{\bf W}^{\ast }(\omega ).  \label{Alphasol1.eq}
\end{equation}
In this result all the terms that in general depend on $\omega $ are
explicitly identified. It is also convenient to write the inverse matrix in
terms of its determinant and the adjugate matrix via:

\begin{equation}
(\omega {\bf E}-{\bf \Omega }(\omega ))^{-1}=(\omega {\bf E}-{\bf \Omega }%
(\omega ))^{ADJ}/|\omega {\bf E}-{\bf \Omega }(\omega )|
\end{equation}
and then the solution for $\bbox{\alpha}(\omega )$ becomes:

\begin{eqnarray}
\bbox{\alpha}(\omega ) &=&{\bf -}i\sqrt{\frac{\rho _{c}(\omega )}{\rho
(\omega )}}\frac{1}{(|\omega {\bf E}-{\bf \Omega }(\omega )|-i\pi \rho
_{c}(\omega ){\bf W}^{T}(\omega )(\omega {\bf E}-{\bf \Omega }(\omega
))^{ADJ}{\bf W}^{\ast }(\omega ))}  \nonumber \\
&&\times (\omega {\bf E}-{\bf \Omega }(\omega ))^{ADJ}{\bf W}^{\ast }(\omega
).  \label{Alphasol2.eq}
\end{eqnarray}

The result for the expansion coefficient $\beta (\omega ,\Delta )$ then
follows from Eq.(\ref{beta.def}) and substituting for $\rho _{c}(\omega
)z(\omega )$ from Eq.(\ref{zresult.eq}). After some algebra we find that:

\begin{equation}
\beta (\omega ,\Delta )=-i\frac{1}{\sqrt{\rho (\omega )\rho _{c}(\omega )}}%
\frac{[\delta (\omega -\Delta )+\frac{{\cal P}}{\omega -\Delta }\rho
_{c}(\omega ){\bf W}^{T}(\omega )(\omega {\bf E}-{\bf \Omega }(\omega ))^{-1}%
{\bf W}^{\ast }(\omega )]}{(1-i\pi \rho _{c}(\omega ){\bf W}^{T}(\omega
)(\omega {\bf E}-{\bf \Omega }(\omega ))^{-1}{\bf W}^{\ast }(\omega ))}
\label{betaresult.eq}
\end{equation}
We see that the solutions for the $\alpha _{i}(\omega )$ and $\beta (\omega
,\Delta )$ only involve the various coupling constants and the mode
densities.

\subsubsection{Coupling constants and reservoir structure function}

Introducing the column matrix $\bbox{\lambda}_{k}\equiv \{\lambda
_{k1},\lambda _{k2},\lambda _{k3},...\}^{\top }$ the expression (\ref
{truemodecouplingconst}) for the coupling constant $g^{k}(\omega )$ can be
written as:

\begin{eqnarray}
g^{k}(\omega ) &=&-i\sqrt{\frac{\rho _{c}(\omega )}{\rho (\omega )}}\frac{1}{%
(|\omega {\bf E}-{\bf \Omega }(\omega )|-i\pi \rho _{c}(\omega ){\bf W}%
^{T}(\omega )(\omega {\bf E}-{\bf \Omega }(\omega ))^{ADJ}{\bf W}^{\ast
}(\omega ))}  \nonumber \\
&&\times \bbox{\lambda}_{k}^{T}(\omega {\bf E}-{\bf \Omega }(\omega ))^{ADJ}%
{\bf W}^{\ast }(\omega )  \label{truecoupresult1} \\
&=&-i\sqrt{\frac{\rho _{c}(\omega )}{\rho (\omega )}}\frac{%
Q_{n-1}^{k}(\omega )}{P_{n}(\omega )}  \label{truecoupresult2}
\end{eqnarray}
where the functions $P_{n}(\omega )$ and $Q_{n-1}^{k}(\omega )$ are defined
by:

\begin{eqnarray}
P_{n}(\omega ) &=&|\omega {\bf E}-{\bf \Omega }(\omega )|-i\pi \rho
_{c}(\omega ){\bf W}^{T}(\omega )(\omega {\bf E}-{\bf \Omega }(\omega
))^{ADJ}{\bf W}^{\ast }(\omega )  \nonumber \\
&=&|\omega {\bf E}-{\bf \Omega }(\omega )|-i\pi \rho _{c}(\omega
)\sum_{ij}W_{i}(\omega )(\omega {\bf E}-{\bf \Omega }(\omega
))_{ij}^{ADJ}W_{j}^{\ast }(\omega )  \label{Pn.def} \\
Q_{n-1}^{k}(\omega ) &=&\bbox{\lambda}_{k}^{T}(\omega {\bf E}-{\bf \Omega }%
(\omega ))^{ADJ}{\bf W}^{\ast }(\omega )  \nonumber \\
&=&\sum_{ij}\lambda _{ki}(\omega {\bf E}-{\bf \Omega }(\omega
))_{ij}^{ADJ}W_{j}^{\ast }(\omega ).  \label{Qn.def}
\end{eqnarray}
In the case where the $\omega $ dependence of the quantities $\rho
_{c}(\omega ),F_{ij}(\omega )$ and $W_{i}(\omega )$ can be ignored, $%
P_{n}(\omega )$ and $Q_{n-1}(\omega )$ would be polynomials in $\omega $ of
degrees $n$ and $n-1$ respectively, as will be seen in Section \ref{SResults}%
.

The reservoir structure function can then be expressed as ($C=1)$:

\begin{equation}
D^{k}(\omega )=\rho _{c}(\omega )\frac{|Q_{n-1}^{k}(\omega )|^{2}}{%
|P_{n}(\omega )|^{2}}  \label{reservfn6}
\end{equation}
where we note the cancellation of the true mode density $\rho (\omega )$ and
the proportionality to the quasi continuum mode density $\rho _{c}(\omega )$%
. The significance of the $\rho (\omega )$ cancellation will be discussed in
Section \ref{SubSectInverseDiag}. There is however further dependence on the
quasi continuum mode density within the function $P_{n}(\omega )$, as can be
seen from Eq.(\ref{Pn.def}). The role of this dependence will be discussed
in Section \ref{SResults} when we have obtained expressions for the
reservoir structure function for specific cases.

To sum up: if we are given the Hamiltonian in the quasi mode form Eq.(\ref
{quasifieldham}), we can obtain the true mode operators (\ref{A.def}) which
satisfy the eigenoperator condition Eq.(\ref{true.def}). The coefficients $%
\alpha _{i}(\omega )$ are found by solving ${\bf m}\bbox{\alpha}=0$, Eq.\ (%
\ref{m1}); the function $z(\omega )$ occuring in ${\bf m}$ is obtained from
Eq.(\ref{m1}) and given by Eq.(\ref{zresult.eq}). The solutions for $\alpha
_{i}(\omega )$ are scaled in accordance with Eq.(\ref{anorm1}) and the
normalisation for the quantity $\sum_{i}W_{i}(\omega )\alpha _{i}(\omega )$
is given in Eqs.(\ref{anorm3}), (\ref{anorm6}). The normalised solutions for 
$\alpha _{i}(\omega )$ are obtained as Eqs.(\ref{Alphasol1.eq}) or (\ref
{Alphasol2.eq}). The coefficients $\beta (\omega ,\Delta )$ are then found
from Eq.(\ref{beta.def}) and the result given in Eq.(\ref{betaresult.eq}).
The true mode coupling constant $g^{k}(\omega )$ and the reservoir structure
function $D^{k}(\omega )$ are obtained as Eqs.(\ref{truecoupresult2})\ and (%
\ref{reservfn6}). These results involve the functions $P_{n}(\omega )$ and $%
Q_{n-1}^{k}(\omega )$ defined in Eqs.(\ref{Pn.def}) and (\ref{Qn.def}). The
results depend on the quasi continuum mode density $\rho _{c}$ as well as on
the various coupling constants and angular frequencies. It should be noted
that a unique expression has been obtained for $z(\omega )$, and hence for
the $\alpha _{i}(\omega )$ and $\beta (\omega ,\Delta )$, even though the
determinental equation $|{\bf m|=}0$ might appear to give anything up to $n$
solutions, where $n$ is the number of discrete quasi modes. This feature is
due to the specific form of the matrix ${\bf m}$ that is involved. The
overall process amounts to a {\em diagonalization} because the EM field
Hamiltonian in the non-diagonal quasi mode form is now replaced by the
diagonal true mode form given by Eq.(\ref{truefieldham}).

\subsection{Inverse diagonalization: undressing the true mode operators}

\label{SubSectInverseDiag}

We can also proceed in the opposite direction from Fano diagonalization:
that is, we can also find the quasi mode operators $\hat{a}_{i}$ and $\hat{b}%
(\Delta )$ in terms of the true mode operators $\hat{A}(\omega )$. In
general (\cite{Barnett88,Dalton99a}) the quasi mode annihilation operators $%
\hat{a}_{i}$ and $\hat{b}(\Delta )$ can also be expressed as linear
combinations of the true mode annihilation operators $\hat{A}(\omega ) $ in
the form: 
\begin{eqnarray}
\hat{a}_{i} &=&\int d\omega \,\rho (\omega )\gamma _{i}(\omega )\hat{A}%
(\omega )  \nonumber \\
\hat{b}(\Delta ) &=&\int d\omega \,\rho (\omega )\delta (\Delta ,\omega )%
\hat{A}(\omega )  \label{ab_try}
\end{eqnarray}
where the functions $\gamma _{i}(\omega )$ and $\delta (\Delta ,\omega )$
have to be determined. These can be obtained in terms of the $\alpha
_{i}(\omega )$ and $\beta (\omega ,\Delta )$ by evaluating the commutators $[%
\hat{A}(\omega ),\hat{a}_{i}^{\dagger }]$ and $[\hat{A}(\omega ),\hat{b}%
(\Delta )^{\dagger }]$ using the basic commutation rules Eqs.(\ref{truecomm}%
), (\ref{quasicomm}). For the first commutator: on substituting for $\hat{A}%
(\omega )$ from Eq.(\ref{A.def}) we obtain $\alpha _{i}(\omega )$, on the
other hand, substituting instead for $\hat{a}_{i}$ from Eq.(\ref{ab_try})
gives $\gamma _{i}^{\ast }(\omega )$, and hence $\alpha _{i}=\gamma
_{i}^{\ast }$. Carrying out a similar process for the second commutator
gives the result $\beta =\delta ^{\ast \text{ }}$and thus: 
\begin{eqnarray}
\hat{a}_{i} &=&\int d\omega \,\rho (\omega )\alpha _{i}^{\ast }(\omega )\hat{%
A}(\omega )  \nonumber \\
\hat{b}(\Delta ) &=&\int d\omega \,\rho (\omega )\beta ^{\ast }(\Delta
,\omega )\hat{A}(\omega )  \label{ab}
\end{eqnarray}
As has been already described in Section \ref{SubSDescription}, the first of
these two equations enables us to relate the two descriptions of the
atom-field interaction given in Eqs.(\ref{quasiatomfieldintn}) and (\ref
{trueatomfieldintn}). Ultimately, the key expression we have obtained in Eq.(%
\ref{truemodecouplingconst}) for the atom-true mode coupling constant rests
on this result. As we will see in Section \ref{SResults}, this enables us to
relate pseudomodes to the discrete quasi modes.

As a final check of the detailed expressions, in Appendix \ref{h=adaga} we
start with the field Hamiltonian in the quasi mode form Eq.(\ref
{quasifieldham}), then substitute our solutions for $\alpha _{i}(\omega )$
and $\beta (\omega ,\Delta )$ into the expressions for $\hat{a}_{i}$ and $%
\hat{b}(\Delta )$ given in Eqs.(\ref{ab}). On evaluating the result, the
Hamiltonian in the true mode form Eq.(\ref{truefieldham}) is obtained---as
required for consistency.

It has already been noted in Section \ref{SubDiagonalisation} that the final
expression for the reservoir structure function $D^{k}(\omega )$ in terms of
quasimode quantities is independent of the true mode density $\rho (\omega )$%
. Also, we have found no equation that actually gives an expression for $%
\rho (\omega )$ in terms of the quasi mode quantities, including the
continuum quasi mode density $\rho _{c}(\Delta )$ - a somewhat surprising
result. The true mode density therefore does not play an important role in
the quasimode theory. The reason for this is not that hard to find, however.
The theory can be recast with {\it both }the $\rho (\omega )$ and $\rho
_{c}(\Delta )$ factors incorporated into the various operators and coupling
constants. In Appendix \ref{App.Scaling} we show that $\rho (\omega )$ and $%
\rho _{c}(\Delta )$ can be scaled away to unity. For example, from Eqs.(\ref
{Alphasol1.eq}), (\ref{betaresult.eq}) and (\ref{A.def}) we see that the
true mode annihilation operator is proportional to $1/\sqrt{\rho (\omega )}$%
, the other (operator) factor only depending on quasi mode quantities. Hence
(as in Appendix \ref{App.Scaling}) we may scale away the $\rho (\omega )$
dependence via the substitution:

\begin{equation}
\hat{A}(\omega )=\frac{\hat{A}^{(s)}(\omega )}{\sqrt{\rho (\omega )}}
\end{equation}
where $\hat{A}^{(s)}(\omega )$ is independent of $\rho $. If this
substitution is made then the field Hamiltonian is given by:

\begin{equation}
\hat{H}_{F}=\int d\omega \,\hbar \omega \hat{A}^{(s)\dagger }(\omega )\hat{A}%
^{(s)}(\omega )
\end{equation}
without any $\rho (\omega )$ term.

\section{ Applications}

\label{SResults}

\subsection{Case of a single quasi mode}

For this case no coupling constant between discrete quasi modes is present
and we may easily allow for a non zero shift matrix element $F_{11}$ and for
non constant $W_{i}(\Delta )$. Noting that $(\omega {\bf E}-{\bf \Omega }%
(\omega ))^{ADJ}=1$ and $|\omega {\bf E}-{\bf \Omega }(\omega )|=\omega
-\omega _{1}-F_{11}(\omega )$, a simple evaluation of Eqs.(\ref{zresult.eq}%
), (\ref{Alphasol1.eq}) and (\ref{truecoupresult2}) gives the following
results: 
\begin{equation}
\rho _{c}(\omega )z(\omega )=\frac{\omega -\omega _{1}-F_{11}(\omega )}{%
|W_{1}(\omega )|^{2}}
\end{equation}
\begin{equation}
\alpha _{1}(\omega )=-i\sqrt{\frac{\rho _{c}(\omega )}{\rho (\omega )}}\frac{%
W_{1}(\omega )^{\ast }}{\omega -\omega _{1}-F_{11}(\omega )-i\pi \rho
_{c}(\omega )|W_{1}(\omega )|^{2}}.
\end{equation}

\[
g^{k}(\omega )=\lambda _{k1}\alpha _{1}(\omega )=-i\sqrt{\frac{\rho
_{c}(\omega )}{\rho (\omega )}}\frac{\lambda _{k1}W_{1}(\omega )^{\ast }}{%
\omega -\omega _{1}-F_{11}(\omega )-i\pi \rho _{c}(\omega )|W_{1}(\omega
)|^{2}} 
\]

In terms of a frequency shift $\Delta \omega _{1}$ and half-width $\frac{%
\Gamma }{2}$ defined as: 
\begin{equation}
\Delta \omega _{1}(\omega )=F_{11}(\omega )
\end{equation}
\begin{equation}
\frac{\Gamma (\omega )}{2}=\pi \rho _{c}(\omega )|W_{1}(\omega )|^{2}
\label{Decay1.eq}
\end{equation}
the reservoir structure function (see Eq.(\ref{reservfn2})) for the
situation where only a single atomic transition $k$ is involved is then
found to be $(C=1)$:

\begin{equation}
D^{k}(\omega )=\frac{|\lambda _{k1}|^{2}\cdot \Gamma (\omega )/2\pi }{%
(\omega -\omega _{1}-\Delta \omega _{1}(\omega ))^{2}+\Gamma (\omega )^{2}/4}
\label{reservfn10}
\end{equation}
In the situation where the quasi mode density $\rho _{c}(\Delta )$ and the
coupling constant $W_{1}(\Delta )$ are slowly varying functions of $\Delta $%
, these quantities can be approximated as constants in the expressions for
the frequency shift and width. The reservoir structure function is then a
Lorentzian shape with a single pole in the lower-half plane at $\omega
_{1}+\Delta \omega _{1}-i\Gamma /2$ corresponding to a single pseudomode.
Thus the single discrete quasi mode is associated with a single pseudomode,
whose position $z_{1}$ is given by $\omega _{1}+\Delta \omega _{1}-i\Gamma
/2 $ in terms of quasi mode quantities.

\subsection{Case of zero discrete quasi mode-quasi mode coupling and flat
reservoir coupling constants}

The theory becomes rather simpler if there is no coupling between the
discrete quasi modes, that is:

\begin{equation}
v_{ij}\Rightarrow 0
\end{equation}
This could be in fact arranged by pre-diagonalising the part of the
Hamiltonian $\hat{H}_{F}$ that only involves the discrete quasi mode
operators. Thus we write 
\begin{equation}
\sum_{i}\hbar \omega _{i}\hat{a}_{i}^{\dagger }\hat{a}_{i}+\sum_{i\neq
j}\hbar v_{ij}\hat{a}_{i}^{\dagger }\hat{a}_{j}
\end{equation}
in the form 
\begin{equation}
\sum_{i}\hbar \xi _{i}\hat{c}_{i}^{\dagger }\hat{c}_{i}
\end{equation}
via the transformation 
\begin{equation}
\hat{c}_{i}=\sum_{j}U_{ij}\hat{a}_{j}.
\end{equation}
where $U$ is unitary. The last equation can be inverted to give the $\hat{a}%
_{i}$ in terms of the $\hat{c}_{i}$ and the result substituted in other
parts of $\hat{H}_{F}$ (Eq.(\ref{quasifieldham})) and $\hat{H}_{AF}$ (Eq.(%
\ref{quasiatomfieldintn})). The original coupling constants $\lambda _{ki}$
and $W_{i}(\Delta )$ would be replaced by new coupling constants via
suitable linear combinations involving the matrix $U$, and these generally
would have similar properties (e.g., flatness) as the original ones.

The idea of replacing the structured reservoir of true modes by quasi modes,
in which the continuum quasi modes constitute a flat reservoir, implies that
the discrete-continuum quasi mode coupling constants $W_{i}(\Delta )$ and
the quasi continuum mode density $\rho _{c}(\Delta )$ are slowly varying
functions of $\Delta $. This results in the shift matrix $F_{ij}$ elements
being small, so it would be appropriate to examine the case where they are
ignored, that is: 
\begin{equation}
F_{ij}\Longrightarrow 0
\end{equation}
with both $\rho _{c}$ and the $W_{i}$ are assumed constant.

For the case $v_{ij}=0,F_{ij}=0,\rho _{c}(\Delta )=\rho _{c}$ and $%
W_{i}(\Delta )=$ $W_{i}$ (constants) the quantities involved in the inverse
of the matrix $\omega {\bf E}-{\bf \Omega }(\omega )$ are:

\begin{eqnarray}
|\omega {\bf E}-{\bf \Omega }(\omega )| &=&(\omega -\omega _{1})(\omega
-\omega _{2})...(\omega -\omega _{n})  \nonumber \\
(\omega {\bf E}-{\bf \Omega }(\omega ))_{ij}^{ADJ} &=&(\omega -\omega
_{1})(\omega -\omega _{2})...(\omega -\omega _{i-1})(\omega -\omega
_{i+1})...(\omega -\omega _{n})\delta _{ij}.
\end{eqnarray}
A straightforward application of Eqs.(\ref{zresult.eq}) and (\ref
{Alphasol2.eq}) leads to the simple results: 
\begin{equation}
\rho _{c}z(\omega )=\left\{ \sum_{i}\frac{|W_{i}|^{2}}{\omega -\omega _{i}}%
\right\}^{-1}
\end{equation}
\begin{equation}
\alpha _{i}(\omega )=-i\sqrt{\frac{\rho _{c}}{\rho (\omega )}}W_{i}^{\ast }%
\frac{(\omega -\omega _{1})(\omega -\omega _{2})...(\omega -\omega
_{i-1})(\omega -\omega _{i+1})...(\omega -\omega _{n})}{P_{n}(\omega )}
\label{alphai}
\end{equation}
where the function $P_{n}(\omega )$, (which is defined in Eq.(\ref{Pn.def}))
is now a polynomial of degree $n$, whose roots are designated as $\xi _{i}$.
It is now given by: 
\begin{eqnarray}
P_{n}(\omega ) &=&(\omega -\omega _{1})(\omega -\omega _{2})...(\omega
-\omega _{n})  \nonumber \\
&&-i\pi \rho _{c}\sum_{j}|W_{j}|^{2}(\omega -\omega _{1})...(\omega -\omega
_{j-1})(\omega -\omega _{j+1})...(\omega -\omega _{n})  \nonumber \\
&=&(\omega -\xi _{1})(\omega -\xi _{2})...(\omega -\xi _{n}).
\end{eqnarray}

For the true mode coupling constants $g^{k}(\omega )$, the general result in
Eq.(\ref{truecoupresult2}) can be applied to give: 
\begin{equation}
g^{k}(\omega )=-i\sqrt{\frac{\rho _{c}}{\rho (\omega )}}\frac{%
Q_{n-1}^{k}(\omega )}{(\omega -\xi _{1})(\omega -\xi _{2})...(\omega -\xi
_{n})}  \label{truecoupresult}
\end{equation}
where the function $Q_{n-1}^{k}(\omega )$ (which is defined in Eq.(\ref
{Qn.def})) is now a polynomial of order $n-1$, whose roots are designated as 
$\theta _{i}$. It is now given by:

\begin{eqnarray}
Q_{n-1}^{k}(\omega ) &=&\sum_{i}\lambda _{ki}W_{i}^{\ast }(\omega -\omega
_{1})(\omega -\omega _{2})...(\omega -\omega _{i-1})(\omega -\omega
_{i+1})...(\omega -\omega _{n})  \label{Qn.eq} \\
&=&S_{k}(\omega -\theta _{1})(\omega -\theta _{2})...(\omega -\theta _{n-1}),
\end{eqnarray}
where $S_{k}$ is a strength factor defined as:

\begin{equation}
S_{k}=\sum_{i}\lambda _{ki}W_{i}^{\ast }  \label{strength.eq}
\end{equation}

The reservoir structure function $D^{k}(\omega )$ (see Eq.(\ref{reservfn6}))
for the $k$ transition is then given by $(C=1)$:

\begin{equation}
D^{k}(\omega )=\rho _{c}|S_{k}|^{2}\frac{|(\omega -\theta _{1})(\omega
-\theta _{2})...(\omega -\theta _{n-1})|^{2}}{|(\omega -\xi _{1})(\omega
-\xi _{2})...(\omega -\xi _{n})|^{2}}.
\end{equation}

Since products of the form $(\omega -\xi )(\omega -\xi ^{\ast })$ can be
written as $(\omega -%
%TCIMACRO{\func{Re}}%
%BeginExpansion
\mathop{\rm Re}%
%EndExpansion
\xi )^{2}+(%
%TCIMACRO{\func{Im}}%
%BeginExpansion
\mathop{\rm Im}%
%EndExpansion
\xi )^{2}$, the behaviour of the reservoir structure function $D^{k}=$ $\rho
(\omega )|g^{k}(\omega )|^{2}$ (see Eq.(\ref{reservfn2})) as a function of $%
\omega $ is now seen to be determined by the product of $n$ Lorentzian
functions associated with $|P_{n}(\omega )|^{2}$ with the modulus squared of
the polynomial of degree $n-1$ given by $|Q_{n-1}^{k}(\omega )|^{2}$. The
quasi continuum mode density merely provides an uninteresting multiplicative
constant, except insofar as it is involved in expressions for the width and
shift factors. In the case where there are $n$ discrete quasi modes, then
irrespective of the location of the roots $\xi _{i}$ of the polynomial
equation $P_{n}(\omega )=0$, the reservoir structure function $D^{k}(\omega
) $ for a single quantum excitation has $n$ poles in the lower half plane,
each corresponding to either $\xi _{i}$ or $\xi _{i}^{\ast }$. As there are $%
n$ roots when $n$ discrete quasimodes are present, we see that each discrete
quasi mode corresponds to one of the $n$ pseudomodes, whose position $z_{i%
\text{ }}$is equal to $\xi _{i}$ or to $\xi _{i}^{\ast }$. Thus, for the
case here where the coupling constants and the quasi continuum mode density
are independent of frequency, the feature that leads to a pseudomode is the
presence of a discrete quasi mode.

\subsection{Case of two discrete quasi modes}

The results in the previous subsection can be conveniently illustrated for
the case of two discrete quasi modes. For simplicity we will again restrict
the treatment to the situation where $v_{12}=0,F_{ij}=0,\rho _{c}(\Delta
)=\rho _{c}$ and $W_{i}(\Delta )=$ $W_{i}$ (constants), and just consider a
two-level atom, so only two coupling constants $\lambda _{1},\lambda _{2}$
are involved. In this case the atom-true mode coupling constant can be
obtained from Eq.(\ref{truecoupresult}) and is:

\begin{equation}
g(\omega )=-i\sqrt{\frac{\rho _{c}}{\rho (\omega )}}\frac{(\lambda
_{1}W_{1}^{\ast }+\lambda _{2}W_{2}^{\ast })(\omega -\omega _{0})}{(\omega
-\xi _{1})(\omega -\xi _{2})}
\end{equation}
where $\omega _{0}$ and the roots $\xi _{1,2}$ of $P_{2}(\omega )=0$ are
given by:

\begin{equation}
\omega _{0}=\frac{\lambda _{2}W_{2}^{\ast }}{(\lambda _{1}W_{1}^{\ast
}+\lambda _{2}W_{2}^{\ast })}\omega _{1}+\frac{\lambda _{1}W_{1}^{\ast }}{%
(\lambda _{1}W_{1}^{\ast }+\lambda _{2}W_{2}^{\ast })}\omega _{2}
\end{equation}
and

\begin{eqnarray}
\xi _{1,2} &=&\frac{1}{2}\{(\omega _{1}+\omega _{2})+i\pi \rho
_{c}(|W_{1}|^{2}+|W_{2}|^{2})\}  \nonumber \\
&&\pm \frac{1}{2}\sqrt{\{(\omega _{1}-\omega _{2})+i\pi \rho
_{c}(|W_{1}|^{2}-|W_{2}|^{2})\}^{2}-4\pi ^{2}\rho
_{c}^{2}|W_{1}|^{2}|W_{2}|^{2}}
\end{eqnarray}
It will also be useful to introduce widths $\Gamma _{i}$ defined by:

\begin{equation}
\Gamma _{i}=2\pi \rho _{c}|W_{i}|^{2}  \label{Decay3.eq}
\end{equation}
and which can be later identified (see Section \ref{SMarkov}) as the
discrete quasi mode decay rates (Eq.(\ref{Decay2.eq})). These results will
be now examined for special subcases.

\subsubsection{Special subcase: Equal quasi mode frequencies}

In this case we choose:

\begin{equation}
\omega _{1}=\omega _{2}=\omega _{C}
\end{equation}
and find that:

\begin{eqnarray}
\omega _{0} &=&\omega _{C}  \nonumber \\
\xi _{1,2} &=&\omega _{C},\omega _{C}+i\pi \rho _{c}(|W_{1}|^{2}+|W_{2}|^{2})
\end{eqnarray}
giving for the atom-true mode coupling constant:

\begin{equation}
g(\omega )=-i\sqrt{\frac{\rho _{c}}{\rho (\omega )}}\frac{(\lambda
_{1}W_{1}^{\ast }+\lambda _{2}W_{2}^{\ast })}{\omega -\omega _{c}+i\pi \rho
_{c}(|W_{1}|^{2}+|W_{2}|^{2})},
\end{equation}
and for the reservoir structure function:

\begin{equation}
D(\omega )=\rho _{c}\frac{|\lambda _{1}W_{1}^{\ast }+\lambda _{2}W_{2}^{\ast
}|^{2}}{(\omega -\omega _{c})^{2}+([\Gamma _{1}+\Gamma _{2}]/2)^{2}}
\label{reservfn8}
\end{equation}

This corresponds to a single pole in the lower half plane for the reservoir
structure function (see Eq.(\ref{reservfn2})) and thus only results in a 
{\it single} pseudomode, albeit for a case of {\it two} degenerate discrete
quasi modes.

\subsubsection{Special subcase: Equal quasi mode reservoir coupling constants%
}

In this case we choose:

\begin{equation}
W_{1}=W_{2}=W
\end{equation}
and find that:

\begin{eqnarray}
\omega _{0} &=&\omega _{C}+\Delta \omega _{C}  \nonumber \\
\omega _{C} &=&\frac{1}{2}(\omega _{1}+\omega _{2})  \nonumber \\
\Delta \omega _{C} &=&\frac{(\lambda _{1}-\lambda _{2})}{2(\lambda
_{1}+\lambda _{2})}(\omega _{2}-\omega _{1})  \nonumber \\
\xi _{1,2} &=&\frac{1}{2}(\omega _{1}+\omega _{2})+i\pi \rho _{c}|W|^{2}\pm 
\frac{1}{2}\sqrt[2]{(\omega _{1}-\omega _{2})^{2}-4\pi ^{2}\rho
_{c}^{2}|W|^{4}}
\end{eqnarray}
Here $\omega _{0}$ has been written in terms of the quasi modes centre
frequency $\omega _{C}$ and a frequency shift $\Delta \omega _{C}$ depending
on the difference between the two atom-discrete quasi modes coupling
constants $\lambda _{i}$ and the discrete quasi modes detuning. There are
now two regimes depending on the relative size of the discrete quasi modes
separation $|\omega _{1}-\omega _{2}|$ compared to the square root of the
quasi continuum mode density $\sqrt[2]{\rho _{c}}$ times the reservoir
coupling constant $W$. Equivalently, the regimes depend on the relative size
of the separation $|\omega _{1}-\omega _{2}|$ compared to the width factor
(decay rate) $\Gamma =\Gamma _{1}=\Gamma _{2}=2\pi \rho _{c}|W|^{2}.$

\paragraph{Regime 1: Large separation $|\protect\omega _{1}-\protect\omega
_{2}|>\Gamma $}

Adopting the convention that $\omega _{1}<\omega _{2}$ we can write

\begin{equation}
\frac{1}{2}\sqrt{(\omega _{1}-\omega _{2})^{2}-4\pi ^{2}\rho _{c}^{2}|W|^{4}}%
=\frac{1}{2}(\omega _{2}-\omega _{1})-\Delta \omega _{R}
\end{equation}
where $\Delta \omega _{R}$ is a reservoir induced frequency shift. The
atom-true mode coupling constant now becomes:

\begin{equation}
g(\omega )=-i\sqrt{\frac{\rho _{c}}{\rho (\omega )}}\frac{(\lambda
_{1}+\lambda _{2})W^{\ast }(\omega -\omega _{C}-\Delta \omega _{C})}{(\omega
-\omega _{2}+\Delta \omega _{R}-i\pi \rho _{c}|W|^{2})(\omega -\omega
_{1}-\Delta \omega _{R}-i\pi \rho _{c}|W|^{2})}
\end{equation}
and the reservoir structure function is then:

\begin{equation}
D(\omega )=\frac{|\lambda _{1}+\lambda _{2}|^{2}(\Gamma /2\pi )(\omega
-\omega _{C}-\Delta \omega _{C})^{2}}{((\omega -\omega _{2}+\Delta \omega
_{R})^{2}+\Gamma ^{2}/4)((\omega -\omega _{1}-\Delta \omega _{R})^{2}+\Gamma
^{2}/4)}  \label{reservfn9}
\end{equation}

The reservoir structure function $D$ (see Eq.(\ref{reservfn9})) will be zero
at the shifted centre frequency $\omega _{C}+\Delta \omega _{C}$. There are
two poles in the lower half plane leading to Lorentzian factors centred at
frequencies $\omega _{2}-\Delta \omega _{R}$ and $\omega _{1}+\Delta \omega
_{R}$ and which have equal widths $2\pi \rho _{c}|W|^{2}$. We note that the
effect of the coupling to the reservoir is to decrease the effective
discrete quasi modes separation by $2\Delta \omega _{R}$.

\paragraph{Regime 2: Small separation $|\protect\omega _{1}-\protect\omega
_{2}|<\Gamma $}

We now write

\begin{equation}
\frac{1}{2}\sqrt{4\pi ^{2}\rho _{c}^{2}|W|^{4}-(\omega _{1}-\omega _{2})^{2}}%
=\pi \rho _{c}|W|^{2}(1-\Delta f_{\Gamma })
\end{equation}
where $\Delta f_{\Gamma }$ is a fractional change in width factors
associated with discrete quasi mode separation. The atom-true mode coupling
constant now becomes:

\begin{equation}
g(\omega )=-i\sqrt{\frac{\rho _{c}}{\rho (\omega )}}\frac{(\lambda
_{1}+\lambda _{2})W^{\ast }(\omega -\omega _{C}-\Delta \omega _{C})}{(\omega
-\omega _{C}-2i\pi \rho _{c}|W|^{2}(1-\frac{1}{2}\Delta f_{\Gamma }))(\omega
-\omega _{C}-i\pi \rho _{c}|W|^{2}\Delta f_{\Gamma })},  \label{atomtrueccB2}
\end{equation}
and the reservoir structure function is:

\begin{equation}
D(\omega )=\frac{|\lambda _{1}+\lambda _{2}|^{2}(\Gamma /2\pi )(\omega
-\omega _{C}-\Delta \omega _{C})^{2}}{((\omega -\omega _{C})^{2}+\Gamma
^{2}(1-\frac{1}{2}\Delta f_{\Gamma })^{2})((\omega -\omega _{C})^{2}+\Gamma
^{2}(\frac{1}{2}\Delta f_{\Gamma })^{2})}  \label{reservfn7}
\end{equation}

The reservoir structure function $D$ (see Eq.(\ref{reservfn7})) will again
be zero at the shifted centre frequency $\omega _{C}+\Delta \omega _{C}$.
There are two poles in the lower half plane leading to Lorentzian factors
both centred at the same frequency $\omega _{C}$, but which have unequal
widths $2\pi \rho _{c}|W|^{2}(1-\frac{1}{2}\Delta f_{\Gamma })$ and $\pi
\rho _{c}|W|^{2}\Delta f_{\Gamma }$. If $\Delta f_{\Gamma }\ll 1$ one width
is much smaller than the other.

In their work on superradiance in a photonic band gap material Bay et al 
\cite{Bay98} assume as a model for the mode density a so-called Fano profile
of the form:

\begin{equation}
\rho (\omega )=\frac{f(\omega -\omega _{C}-q)^{2}}{((\omega -\omega
_{C})^{2}+(\frac{1}{2}\kappa )^{2})((\omega -\omega _{C})^{2}+(\frac{1}{2}%
\gamma )^{2})}
\end{equation}
with the two-level atom coupling constant $g(\omega )$ given by a slowly
varying function proportional to $\sqrt{\omega }$. It is interesting to note
that the reservoir structure function related to their theory is of the same
form as that obtained here from Eq.(\ref{reservfn7}) if the following
identifications are made:

\begin{eqnarray}
q &\longrightarrow &-\Delta \omega _{C}  \nonumber \\
\frac{1}{2}\kappa &\rightarrow &2\pi \rho _{c}|W|^{2}(1-\frac{1}{2}\Delta
f_{\Gamma })  \nonumber \\
\frac{1}{2}\gamma &\rightarrow &\pi \rho _{c}|W|^{2}\Delta f_{\Gamma }
\end{eqnarray}
For situations such as atomic systems coupled to the field in high Q
cavities the physics is different of course, with the resonant behaviour in
the reservoir structure function being due to the atom-true mode coupling
constants rather than the reservoir mode density (which we assume is slowly
varying). Nevertheless, our two discrete quasi mode model---with equal
reservoir coupling constants $W$ that are large compared to the discrete
quasi modes detuning $|\omega _{1}-\omega _{2}|$---does provide an {\it %
equivalent} physical model for the photonic band gap case that Bay et al
treated, the lack of which was commented on in the review by Lambropoulos et
al \cite{Lambropoulos00}.

The band gap case was also treated as a specific example by Garraway \cite
{Garraway97} in the original pseudomode theory paper. A model for the
reservoir structure function was assumed in the form of a difference between
two Lorentzians:

\begin{equation}
D(\omega )=w_{1}\frac{\Gamma _{1}}{(\omega -\omega _{C})^{2}+(\frac{1}{2}%
\Gamma _{1})^{2}}-w_{2}\frac{\Gamma _{2}}{(\omega -\omega _{C})^{2}+(\frac{1%
}{2}\Gamma _{2})^{2}}
\end{equation}
where the weights $w_{1},w_{2}$ satisfy $w_{1}-w_{2}=1$. Again, apart from
an overall proportionality constant this same form can be obtained here (see
Eq.(\ref{reservfn7})) for the reservoir structure function $D$ if we choose
the atom-discrete quasi mode coupling constants $\lambda _{1},\lambda _{2}$
to be equal (so that the frequency shift $\Delta \omega _{C}$ is zero):

\begin{eqnarray}
\lambda _{1} &=&\lambda _{2}  \nonumber \\
\Delta \omega _{C} &=&0
\end{eqnarray}
and where the following identifications are made:

\begin{eqnarray}
\frac{1}{2}\Gamma _{1} &\rightarrow &2\pi \rho _{c}|W|^{2}(1-\frac{1}{2}%
\Delta f_{\Gamma })  \nonumber \\
\frac{1}{2}\Gamma _{2} &\rightarrow &\pi \rho _{c}|W|^{2}\Delta f_{\Gamma } 
\nonumber \\
w_{1} &\rightarrow &\frac{1-\frac{1}{2}\Delta f_{\Gamma }}{1-\Delta
f_{\Gamma }}  \nonumber \\
w_{2} &\rightarrow &\frac{\frac{1}{2}\Delta f_{\Gamma }}{1-\Delta f_{\Gamma }%
}.
\end{eqnarray}
As will be seen in the Section \ref{SMarkov} the existence of unusual forms
of the reservoir structure function (such as the presence of Lorentzians
with negative weights) does not rule out Markovian master equations applying
to the atom-discrete quasi modes system. Thus, for the situation of single
quantum excitation, where the pseudomodes are always equivalent to discrete
quasi modes, we can always obtain Markovian master equations for
pseudomode-atom system.

\section{Markovian Master Equation for Atom-Discrete Quasi Modes System}

\label{SMarkov}

A key idea for treating the behaviour of a small system coupled to a
structured reservoir is that although the behavior of the small system
itself is non-Markovian, an enlarged system can obtained that exhibits
Markovian dynamics---and which includes the small system, whose dynamics can
be obtained later. In our example of a multilevel atomic system coupled to
the the quantum EM\ field as a structured reservoir, we can proceed as
follows. The overall system of the atom(s) plus quantum EM field is
partitioned into a Markovian system consisting of the atom plus the discrete
quasi modes and a flat reservoir consisting of the continuum quasi modes.
The system Hamiltonian $\hat{H}_{S}$ is: 
\begin{eqnarray}
\hat{H}_{S} &=&\sum_{k}\eta _{k}\hbar \omega _{k}\left( \hat{\sigma}_{k}^{+}%
\hat{\sigma}_{k}^{-}-\hat{\sigma}_{k}^{-}\hat{\sigma}_{k}^{+}\right)
+\sum_{i}\hbar \omega _{i}\hat{a}_{i}^{\dagger }\hat{a}_{i}  \nonumber \\
&&+\sum_{i\neq j}\hbar v_{ij}\hat{a}_{i}^{\dagger }\hat{a}%
_{j}+\sum_{k}\sum_{i}\left( \hbar \lambda _{ki}^{\ast }\hat{a}_{i}\hat{\sigma%
}_{k}^{+}+\mbox{h.c.}\right)
\end{eqnarray}
whilst the reservoir Hamiltonian $\hat{H}_{R}$ is: 
\begin{equation}
\hat{H}_{R}=\int d\Delta \rho _{c}(\Delta )\,\hbar \Delta \hat{b}^{\dagger
}(\Delta )\hat{b}(\Delta )
\end{equation}
and the system-reservoir interaction Hamiltonian $\hat{H}_{{\rm S-R}}$ is: 
\begin{equation}
\hat{H}_{{\rm S-R}}=\sum_{i}\int d\Delta \rho _{c}(\Delta )\left[ \hbar
W_{i}(\Delta )\hat{a}_{i}^{\dagger }\hat{b}(\Delta )+\mbox{h.c.}\right] ,
\end{equation}
so that the total Hamiltonian is still equal to the sum of $\hat{H}_{A},\hat{%
H}_{F}$ and $\hat{H}_{AF}$, given in Eqs.(\ref{atomham}), (\ref
{quasifieldham}) and (\ref{quasiatomfieldintn}). The distinction between the
non-Markovian true mode treatment and the Markovian quasi mode approach is
depicted in Fig. \ref{Figure 2}.

It is of course the slowly varying nature of the coupling constants $%
W_{i}(\Delta )$ and the mode density $\rho _{c}(\Delta )$ which results in a
Markovian master equation for the reduced density operator $\,\hat{\rho}$ of
the atom-discrete quasi modes system. Rather than derive the master equation
for the most general state of the reservoir, we will just consider the
simplest case in which the reservoir of continuum quasi modes are all in the
vacuum state. Again, the coupling constants $W_{i}$ will be assumed constant
so that no shift matrix $F_{ij}$ elements are present. The master equation
is derived via standard proceedures (Born and Markoff approximations) \cite
{Barnett97,DaltonFS99}, which require the evaluation of two-time reservoir
correlation functions in which the required reservoir operators are the
quantities $\int d\Delta \rho _{c}(\Delta )W_{i}(\Delta )\hat{b}(\Delta )$
and their Hermitian adjoints. To obtain Markovian behaviour we require the
quantities $\rho _{c}(\Delta )W_{i}(\Delta )W_{j}^{\ast }(\Delta )$ to be
slowly varying with $\Delta $, so that the reservoir correlation time $\tau
_{c}$ (inversely proportional to the bandwidth of $\rho _{c}(\Delta
)W_{i}(\Delta )W_{j}^{\ast }(\Delta )$) is sufficiently short that the
interaction picture density operator hardly changes during $\tau _{c}$.

The standard procedure then yields the master equation in the Lindblad form: 
\begin{equation}
\frac{d\hat{\rho}}{dt}=\frac{-i}{\hbar }\left[ \hat{H}_{S}\,,\,\hat{\rho}%
\right] \,+\,\sum_{ij}\pi \rho _{c}W_{i}W_{j}^{\ast }\left\{ \left[ \hat{a}%
_{j},\hat{\rho}\hat{a}_{i}^{\dagger }\right] +\left[ \hat{a}_{j}\hat{\rho},%
\hat{a}_{i}^{\dagger }\right] \right\}  \label{LindbladME}
\end{equation}
Direct couplings between the discrete quasi modes involving the $v_{ij}$ are
included in the system Hamiltonian $\hat{H}_{S}$. Radiative processes take
place via the atom-discrete quasi modes interaction also included in $\hat{H}%
_{S}$, though still given as in Eq.(\ref{quasiatomfieldintn}). The loss of
radiative energy to the reservoir is described via the relaxation terms in
the master equation. The diagonal terms where $i=j$ describe the relaxation
of the $i$ th quasi mode in which the decay rate is proportional to $\rho
_{c}|W_{i}|^{2}$. A typical decay rate $\Gamma _{i}$ for the $i$ th discrete
quasi mode into the reservoir of continuum quasi modes will be:

\begin{equation}
\Gamma _{i}=2\pi \rho _{c}|W_{i}|^{2}  \label{Decay2.eq}
\end{equation}
Note that the off-diagonal terms $i\neq j$ involve pairs of discrete quasi
mode operators $\hat{a}_{j}$ and $\hat{a}_{i}^{\dagger }$, so there is also
a type of rotating wave approximation interaction taking place via the
reservoir between these discrete quasi modes, as well as via direct
Hamiltonian coupling involving the $v_{ij}$. The standard criterion for the
validity of the Born-Markoff master equation Eq.(\ref{LindbladME}) is that $%
\Gamma \tau _{c}\ll 1$. Processes involving multiphoton excitation of the
reservoir (such as may occur for excited multilevel atoms) can be studied
using standard master equation methods, thereby enabling multiple excitation
of the structured reservoir to be treated via the quasimode theory.

\section{Non Slowly-Varying Mode Densities and/or Coupling Constants}

\label{SVarying}

The basic model treated in this paper is that of atomic systems coupled to a
set of discrete quasi modes of the EM field, which are in turn coupled to a
continuum set of quasi modes. Although expressions for the true mode
coupling constant and the reservoir structure function have been obtained
for the general case where the quasi mode density $\rho _{c}$ and the
coupling constants $W_{i}$ are not necessarily slowly varying functions of $%
\Delta $ (see Eqs.(\ref{truecoupresult2}) and (\ref{reservfn6})), the
usefulness of the results where this is not the case is somewhat limited. As
indicated in the previous section, the master equation for the atom plus
discrete quasi modes system will no longer be Markovian, so the enlargement
of the system based on adding the discrete quasi modes to produce a
Markovian system fails.

Also, for the non slowly varying $\rho _{c}$ or $W_{i}$ case, we can no
longer link each discrete quasi mode to a pseudo mode. That this is the
situation may be seen both from the general result for the reservoir
structure function (Eq.(\ref{reservfn6})) or the specific result we have
obtained for the case where there is a single discrete quasi mode (Eq.(\ref
{reservfn10})). In the former case, the function $P_{n}(\omega )$ would not
be a polynomial of degree $n$, and therefore could have more than $n$ roots,
leading to more pseudo modes than discrete quasi modes. In the latter case
involving just one discrete quasi mode, even having the mode density $\rho
_{c}(\omega )$ (and hence $\Gamma (\omega )$) represented by a single peaked
function would result in $D(\omega )$ going from a single peaked function to
a triple peaked function, corresponding to three pseudo modes.

However, where $\rho _{c}$ or $W_{i}$ are no longer slowly varying, an
examination of the underlying causes for this variation may suggest
replacing the present atom plus discrete and continuum quasi mode model by a
more elaborate system that better represents the physics of the situation,
but with only slowly varying parameters now involved. Fano diagonalisation
based on such a more elaborate model could produce the desired link up with
the pseudo mode approach and enable a suitable, enlarged system to be
identified which has Markovian behaviour, as well as overcoming the problem
of treating multiple reservoir excitations. One possible elaboration would
be to add a further continuum of quasi modes that are fermionic rather than
bosonic.

\section{ Conclusions}

\label{SConc}

The theory presented above is mainly intended to apply to the important
situation where the reservoir structure is actually due to the presence of a
discrete system of quasi modes which are coupled to other continuum quasi
modes via slowly varying coupling constants. For example, the quantum EM\
field in high Q resonant cavities can be accurately described in terms of
the quasi mode model which has these features, the discrete quasi modes
being the cavity quasi modes (linked to the cavity resonances) with which
the atoms inside the cavity interact, and the continuum quasi modes being
the external modes.

For this situation it has been shown that, for the present case of single
quantum excitations, the pseudo mode method for treating atomic systems
coupled to a structured reservoir of true quantum EM field modes can be
obtained by applying the Fano diagonalisation method to the field described
in an equivalent way as a set of discrete quasi modes together with a set of
continuum quasi modes, whose mode density is assumed to be slowly varying.
The interaction between the discrete and continuum quasi modes is treated in
the rotating-wave approximation assuming slowly varying coupling constants,
and the atomic system is assumed to be only coupled to the discrete quasi
modes. The theory includes the true and continuum quasi mode densities
explicitly.

Expressions for the quasi mode operators $\hat{a}_{i}$ and $\hat{b}(\Delta )$
in terms of the true mode operators $\hat{A}(\omega )$ (and vice versa) have
been found, and explicit forms for the atom-true mode coupling constants
have been obtained and related to the reservoir structure function that
applies in pseudomode theory. We have seen that the feature that leads to a
pseudomode is the presence of a discrete quasi mode. Each discrete quasi
mode corresponds to one of the pseudomodes, whose position $z_{i\text{ }}$in
the lower half complex plane is determined from the roots $\xi _{i}$ of a
polynomial equation depending on the parameters for the quasi mode system.

Although the behavior of the atom itself is non-Markovian, an enlarged
system consisting of the atom plus the discrete quasi modes coupled to a
flat reservoir consisting of the continuum quasi modes exhibits Markovian
dynamics, and the master equation for this enlarged system has been
obtained. Using the quasimode theory, processes involving multiphoton
excitation of the structured reservoir (such as may occur for excited
multilevel atoms) can now be studied using standard master equation methods
applied to the atom-discrete quasi modes system. Furthermore, cases with
unusual forms of the reservoir structure function for single quantum
excitation (for example, containing Lorentzians with negative weights) still
result in Markovian master equations. Since for single quantum excitation
the pseudomodes are equivalent to discrete quasi modes, we can now always
obtain Markovian master equations for pseudomode-atom systems via our
approach.

Although not so useful in such cases, the present theory does lead to
general expressions for the true mode coupling constant and the reservoir
structure function for single quantum excitation. These expressions are
still valid for the general case where the quasi mode density $\rho _{c}$
and the coupling constants $W_{i}$ are no longer slowly varying functions of 
$\Delta $. However, the master equation for the atom plus discrete quasi
modes system will no longer be Markovian, so the enlargement of the system
based on adding the discrete quasi modes to produce a Markovian system
fails. Also, for the non slowly varying $\rho _{c}$ or $W_{i}$ case, we can
no longer link each discrete quasi mode to a pseudo mode---there may be more
pseudo modes than discrete quasi modes. In such cases it would be desirable
to replace the present quasi mode system by a more elaborate quasi mode
system involving only slowly varying quantities, and which better represents
the underlying physical causes of the variation in $W_{i}$ and $\rho _{c}$
that occurs in the present model. This may make possible an extension of the
Fano diagonalisation approach that still links quasi modes with pseudo
modes, and results in a Markovian master equation for the enlarged atom plus
quasi mode system. In such an elaborated system, the disadvantage of the
present pseudo mode treatment in treating multiple excitations of the
structured reservoir could still be removed.

The treatment has been outlined in the case of a multilevel atom coupled to
a structured reservoir of quantum EM\ field modes, but a similar approach
would apply for any fermionic system coupled to a structured reservoir of
bosonic oscillators. Extensions to fermionic reservoirs should be possible
also. At present the treatment is restricted to cases where threshold and
band gap effects are unimportant, but may be applicable to two-dimensional
photonic band gap materials. Further extensions of the treatment to allow
for atomic systems driven by single mode external laser fields are also
possible, with the original atomic system being replaced by the dressed atom.

\section{ Acknowledgments}

The authors are grateful to D. G. Angelakis, J. Wang and A. Imamoglu for
helpful discussions. This work was supported by the United Kingdom
Engineering and Physical Sciences Research Council. SMB thanks the Royal
Society of Edinburgh and the Scottish Executive Education and Lifelong
Learning Department for the award of a Support Research Fellowship.

\appendix

\section{ Atomic Hamiltonian}

\label{App. Atomic Ham}

As an example of writing the atomic Hamiltonian in the form given in Eq.(\ref
{atomham}), consider a three level atom in a V configuration with upper
states$\ |2\rangle ,|1\rangle $ and lower state $|0\rangle $, whose energy
is chosen for convenience to be zero. The atomic transition operators are $%
\hat{\sigma}_{2}^{+}\equiv |2\rangle \langle 0|$ and $\ \hat{\sigma}%
_{1}^{+}\equiv |1\rangle \langle 0|$ for the two optical transitions of
frequencies $\omega _{2}$ and $\omega _{1}$, and $\hat{\sigma}_{3}^{+}\equiv
|2\rangle \langle 1|$ for the Zeeman transition of frequency $\omega
_{2}-\omega _{1}$.

The form given in Eq.(\ref{atomham}) is:

\begin{eqnarray}
\hat{H}_{A} &=&\eta _{1}\hbar \omega _{1}\left( \hat{\sigma}_{1}^{+}\hat{%
\sigma}_{1}^{-}-\hat{\sigma}_{1}^{-}\hat{\sigma}_{1}^{+}\right) +\eta
_{2}\hbar \omega _{2}\left( \hat{\sigma}_{2}^{+}\hat{\sigma}_{2}^{-}-\hat{%
\sigma}_{2}^{-}\hat{\sigma}_{2}^{+}\right) +\eta _{3}\hbar (\omega
_{2}-\omega _{1})\left( \hat{\sigma}_{3}^{+}\hat{\sigma}_{3}^{-}-\hat{\sigma}%
_{3}^{-}\hat{\sigma}_{3}^{+}\right)  \nonumber \\
&=&\eta _{1}\hbar \omega _{1}\left( |1\rangle \langle 1|-|0\rangle \langle
0|\right) +\eta _{2}\hbar \omega _{2}\left( |2\rangle \langle 2|-|0\rangle
\langle 0|\right) +\eta _{3}\hbar (\omega _{2}-\omega _{1})\left( |2\rangle
\langle 2|-|1\rangle \langle 1|\right) .  \label{AH1}
\end{eqnarray}
This expression may also be written in the form

\begin{equation}
\hat{H}_{A}=\hbar \omega _{1}|1\rangle \langle 1|+\hbar \omega _{2}|2\rangle
\langle 2|+\hbar \omega (|0\rangle \langle 0|+|1\rangle \langle 1|+|2\rangle
\langle 2|),  \label{AH2}
\end{equation}
since by equating the coefficients of the three projection operators, we
obtain a set of linear equations for the $\eta _{1},\eta _{2},\eta _{3}$ and 
$\omega $ which are solvable $-$ in fact the solutions are not even unique.
These equations are:

\begin{eqnarray}
\eta _{2}\omega _{2}+\eta _{3}(\omega _{2}-\omega _{1}) &=&\omega _{2}+\omega
\nonumber \\
\eta _{1}\omega _{1}-\eta _{3}(\omega _{2}-\omega _{1}) &=&\omega _{1}+\omega
\nonumber \\
-\eta _{1}\omega _{1}-\eta _{2}\omega _{2} &=&\omega .  \label{AHEQ3}
\end{eqnarray}

Adding these equations and then substituting into the first two gives:

\begin{eqnarray}
\omega &=&-\frac{1}{3}(\omega _{1}+\omega _{2})  \label{AHEQ6} \\
+\eta _{3}(\omega _{2}-\omega _{1}) &=&-\frac{1}{3}\omega _{1}+(\frac{2}{3}%
-\eta _{2})\omega _{2}  \nonumber \\
-\eta _{3}(\omega _{2}-\omega _{1}) &=&(\frac{2}{3}-\eta _{1})\omega _{1}-%
\frac{1}{3}\omega _{2}.  \label{AHEQ4}
\end{eqnarray}
The last two equations do not produce a unique solution for $\eta _{1},\eta
_{2},\eta _{3}$. We can arbitarily choose $\eta _{3}=0$ for the low
frequency transition, and then we find that:

\begin{eqnarray}
\eta _{1} &=&\frac{1}{3}-\frac{1}{3}\frac{(\omega _{2}-\omega _{1})}{\omega
_{1}}  \nonumber \\
\eta _{2} &=&\frac{1}{3}+\frac{1}{3}\frac{(\omega _{2}-\omega _{1})}{\omega
_{2}}.  \label{AHEQ5}
\end{eqnarray}
This gives $\eta _{1}$ $=\eta _{2}=\frac{1}{3}$ for two degenerate optical
frequency transitions.

Comparing the two expressions for $\hat{H}_{A}$ in Eqs.(\ref{AH1}), (\ref
{AH2}), where $\eta _{1},\eta _{2}$ are given by Eq.(\ref{AHEQ5}) (with $%
\eta _{3}$ set to zero) and $\omega $ by Eq.(\ref{AHEQ6}), we see that Eq.(%
\ref{AH2}) gives the atomic energy apart from the constant term $-\frac{1}{3}%
\hbar (\omega _{1}+\omega _{2})$.

\section{ Scaling for Mode Densities $\protect\rho (\protect\omega ),\protect%
\rho _{c}(\Delta )$ Equal to Unity}

\label{App.Scaling}

The equations presented in the first part of Section \ref{SFano} are based
on true and quasi continuum mode densities that are not necessarily equal to
unity. To compare our expressions with those in the Ref.\cite{Barnett97}, we
now set out the scalings needed for the various quantities to give the
Hamiltonians equivalent to $\hat{H}_{F}$ and $\hat{H}_{AF}$ in either true
or quasi mode forms (Eqs.(\ref{truefieldham}), (\ref{quasifieldham}), (\ref
{trueatomfieldintn}) and (\ref{quasiatomfieldintn})) in which the mode
densities $\rho $ and $\rho _{c}$ are made equal to unity. The creation and
anihilation operators are no longer dimensionless, the coupling constants
and angular frequencies do not have dimensions of frequency and the
expansion coefficients are not dimensionless. The scaled quantities
appearing in the Hamiltonians or relationships between annihilation
operators will be denoted with a superscript $^{(s)}$.

The following replacements made to the annihilation and creation operators:

\begin{eqnarray}
\sqrt{\rho _{c}(\Delta )}\hat{b}(\Delta ) &\rightarrow &\hat{b}^{(s)}(\Delta
) \\
\sqrt{\rho _{c}(\Delta )}\hat{b}^{\dagger }(\Delta ) &\rightarrow &\hat{b}%
^{(s)\dagger }(\Delta ) \\
\sqrt{\rho (\omega )}\hat{A}(\omega ) &\rightarrow &\hat{A}^{(s)}(\omega ) \\
\sqrt{\rho (\omega )}\hat{A}^{\dagger }(\omega ) &\rightarrow &\hat{A}%
^{(s)\dagger }(\omega ),
\end{eqnarray}
to the coupling constants:

\begin{equation}
\sqrt{\rho _{c}(\Delta )}W_{i}(\Delta )\rightarrow W_{i}^{(s)}(\Delta )
\end{equation}
and to the expansion coefficients:

\begin{eqnarray}
\sqrt{\rho (\omega )}\alpha _{i}(\omega ) &\rightarrow &\alpha _{i}(\omega
)^{(s)} \\
\sqrt{\rho _{c}(\Delta )\rho (\omega )}\beta (\omega ,\Delta ) &\rightarrow
&\beta ^{(s)}(\omega ,\Delta )
\end{eqnarray}
will give the Hamiltonians equivalent to $\hat{H}_{F}$ and $\hat{H}_{AF}$ in
either true or quasi mode forms (Eqs.(\ref{truefieldham}),(\ref
{quasifieldham}),(\ref{trueatomfieldintn}) and (\ref{quasiatomfieldintn}))
in which the mode densities are put equal to one. In addition the modified
forms of the relationships between true and quasi mode annihilation
operators (Eqs.(\ref{A.def}) and (\ref{ab})) can be obtained in which $\rho $
and $\rho _{c}$ are made equal to unity, as can the revised forms of the
commutation rules. The latter are:

\begin{eqnarray}
\lbrack \hat{b}^{(s)}(\Delta ),\hat{b}^{(s)\dagger }(\Delta ^{\prime })]
&=&\delta (\Delta -\Delta ^{\prime }) \\
\lbrack \hat{A}^{(s)}(\omega ),\hat{A}^{(s)\dagger }(\omega ^{\prime })]
&=&\delta (\omega -\omega ^{\prime }).
\end{eqnarray}
In addition the various equations for the $F_{ij}(\omega ),\alpha
_{i}(\omega ),\beta (\omega ,\Delta ),z(\omega ),g^{k}(\omega ),D^{k}(\omega
)$ now apply with $\rho $ and $\rho _{c}$ put equal to unity. It should be
noted that the quantities $\hat{a}_{i},\omega _{i},v_{ij},\Delta ,\omega $
are not replaced, nor are any of the atomic quantities $\hat{H}_{A},\eta
_{k},\omega _{k},\hat{\sigma}_{k}^{+},\hat{\sigma}_{k}^{-}$ or $\lambda
_{ki} $.

\section{ The Hamiltonian $H_{F}$ in Diagonalised Form}

\label{h=adaga}

We show by starting with the field Hamiltonian in the quasi mode form Eq.(%
\ref{quasifieldham}), substituting the solutions for $\alpha _{i}(\omega )$
and $\beta (\omega ,\Delta )$ into the expressions for $\hat{a}_{i}$ and $%
\hat{b}(\Delta )$ given in Eqs.(\ref{ab}) and then evaluating the result,
that the Hamiltonian in the true mode form Eq.(\ref{truefieldham}) is
obtained. The symmetry conditions $F_{ij}=F_{ji}^{\ast }$ and $\nu _{ij}=\nu
_{ji}^{\ast }$ are used throughout.

Using the expressions for $\hat{a}_{i}$ and $\hat{b}(\Delta )$ given in Eqs.(%
\ref{ab}) the Hamiltonian in the quasi mode form Eq.(\ref{quasifieldham}) is
then given by:

\begin{equation}
\hat{H}_{F}=\hbar \int d\omega \rho (\omega )\,\int d\omega ^{/}\rho (\omega
^{/})\hat{A}^{\dagger }(\omega )\hat{A}(\omega ^{/})I(\omega ,\omega ^{/})
\label{Hstart.eq}
\end{equation}
where the function $I(\omega ,\omega ^{/})$ is:

\begin{eqnarray}
I(\omega ,\omega ^{/}) &=&\sum_{i}\omega _{i}\alpha _{i}(\omega )\alpha
_{i}^{\ast }(\omega ^{/})+\int d\Delta \rho _{c}(\Delta )\Delta \beta
(\Delta ,\omega )\beta ^{\ast }(\Delta ,\omega ^{/})  \nonumber \\
&&+\sum_{ij(i\neq j)}\nu _{ij}\alpha _{i}(\omega )\alpha _{j}^{\ast }(\omega
^{/})+\sum_{i}\int d\Delta \rho _{c}(\Delta )W_{i}(\Delta )\alpha
_{i}(\omega )\beta ^{\ast }(\Delta ,\omega ^{/})  \nonumber \\
&&+\sum_{i}\int d\Delta \rho _{c}(\Delta )W_{i}^{\ast }(\Delta )\alpha
_{i}^{\ast }(\omega ^{/})\beta (\Delta ,\omega )  \label{Iww.eq}
\end{eqnarray}

Substituting for $\beta (\omega ,\Delta )$ in terms of the $\alpha
_{i}(\omega )$ from Eq.(\ref{beta.def}), using the expression (\ref{Fij.def}%
) for $F_{ij}$ and then Eq.(\ref{z1}) for the $\alpha _{i}(\omega )$ we get
for certain contributions within the last two terms in Eq.(\ref{Iww.eq}):

\begin{eqnarray}
\int d\Delta \rho _{c}(\Delta )W_{i}(\Delta )\beta ^{\ast }(\Delta ,\omega
^{/}) &=&-(\omega _{i}-\omega ^{/})\alpha _{i}^{\ast }(\omega
^{/})-\sum_{j(j\neq i)}\nu _{ji}^{\ast }\alpha _{j}^{\ast }(\omega ^{/}) 
\nonumber \\
\int d\Delta \rho _{c}(\Delta )W_{i}^{\ast }(\Delta )\beta (\Delta ,\omega )
&=&-(\omega _{i}-\omega )\alpha _{i}(\omega )-\sum_{j(j\neq i)}\nu
_{ji}\alpha _{j}(\omega )
\end{eqnarray}
leading to

\begin{eqnarray}
\sum_{i}\int d\Delta \rho _{c}(\Delta )W_{i}(\Delta )\alpha _{i}(\omega
)\beta ^{\ast }(\Delta ,\omega ^{/}) &=&-\sum_{i}(\omega _{i}-\omega
^{/})\alpha _{i}(\omega )\alpha _{i}^{\ast }(\omega ^{/})-\sum_{ij(j\neq
i)}\nu _{ji}^{\ast }\alpha _{i}(\omega )\alpha _{j}^{\ast }(\omega ^{/})
\label{t4.eq} \\
\sum_{i}\int d\Delta \rho _{c}(\Delta )W_{i}^{\ast }(\Delta )\alpha
_{i}^{\ast }(\omega ^{/})\beta (\Delta ,\omega ) &=&-\sum_{i}(\omega
_{i}-\omega )\alpha _{i}(\omega )\alpha _{i}^{\ast }(\omega
^{/})-\sum_{ij(j\neq i)}\nu _{ji}\alpha _{j}(\omega )\alpha _{i}^{\ast
}(\omega ^{/})  \label{t5.eq}
\end{eqnarray}

In the second term of Eq.(\ref{Iww.eq}) substitution for $\beta (\omega
,\Delta )$ and $\beta ^{\ast }(\omega ^{/},\Delta )$ in terms of the $\alpha
_{i}(\omega )$ and $\alpha _{j}^{\ast }(\omega ^{/})$from Eq.(\ref{beta.def}%
) and then using the Eqs.(\ref{delta_tech1}) for manipulating principal
integrals and delta functions leads to:

\begin{eqnarray}
\int d\Delta \rho _{c}(\Delta )\Delta \beta (\Delta ,\omega )\beta ^{\ast
}(\Delta ,\omega ^{/}) &=&\sum_{ij}\{[\int d\Delta \rho _{c}(\Delta )\times 
\nonumber \\
&&\times \Delta \frac{{\cal P}}{\omega ^{/}-\omega }\left( \frac{{\cal P}}{%
\omega -\Delta }-\frac{{\cal P}}{\omega ^{/}-\Delta }\right) W_{i}(\Delta
)\alpha _{i}(\omega )W_{j}^{\ast }(\Delta )\alpha _{j}^{\ast }(\omega ^{/})]
\nonumber \\
&&+\pi ^{2}\delta (\omega -\omega ^{/})\rho _{c}(\omega )\omega W_{i}(\omega
)\alpha _{i}(\omega )W_{j}^{\ast }(\omega )\alpha _{j}^{\ast }(\omega ) 
\nonumber \\
&&+\omega ^{/}\frac{{\cal P}}{\omega -\omega ^{/}}\rho _{c}(\omega
^{/})z^{\ast }(\omega ^{/})W_{i}(\omega ^{/})\alpha _{i}(\omega )W_{j}^{\ast
}(\omega ^{/})\alpha _{j}^{\ast }(\omega ^{/})  \nonumber \\
&&+\omega \frac{{\cal P}}{\omega ^{/}-\omega }\rho _{c}(\omega )z(\omega
)W_{i}(\omega )\alpha _{i}(\omega )W_{j}^{\ast }(\omega )\alpha _{j}^{\ast
}(\omega ^{/})  \nonumber \\
&&+\omega \delta (\omega -\omega ^{/})\rho _{c}(\omega )z(\omega )z^{\ast
}(\omega )W_{i}(\omega )\alpha _{i}(\omega )W_{j}^{\ast }(\omega )\alpha
_{j}^{\ast }(\omega )\}
\end{eqnarray}
Then using Eq.(\ref{delta_tech2}) we show that:

\begin{equation}
\Delta \left( \frac{{\cal P}}{\omega -\Delta }-\frac{{\cal P}}{\omega
^{/}-\Delta }\right) =\left( \omega \frac{{\cal P}}{\omega -\Delta }-\omega
^{/}\frac{{\cal P}}{\omega ^{/}-\Delta }\right)
\end{equation}
and following the introduction of the $F_{ij}$ from Eq.(\ref{Fij.def}) we
get:

\begin{eqnarray}
\int d\Delta \rho _{c}(\Delta )\Delta \beta (\Delta ,\omega )\beta ^{\ast
}(\Delta ,\omega ^{/}) &=&\sum_{ij}\{\omega \frac{{\cal P}}{\omega
^{/}-\omega }F_{ji}(\omega )\alpha _{i}(\omega )\alpha _{j}^{\ast }(\omega
^{/})  \nonumber \\
&&-\omega ^{/}\frac{{\cal P}}{\omega ^{/}-\omega }F_{ji}(\omega ^{/})\alpha
_{i}(\omega )\alpha _{j}^{\ast }(\omega ^{/})  \nonumber \\
&&+\omega \frac{{\cal P}}{\omega ^{/}-\omega }\rho _{c}(\omega )z(\omega
)W_{i}(\omega )\alpha _{i}(\omega )W_{j}^{\ast }(\omega )\alpha _{j}^{\ast
}(\omega ^{/})  \nonumber \\
&&-\omega ^{/}\frac{{\cal P}}{\omega ^{/}-\omega }\rho _{c}(\omega
^{/})z^{\ast }(\omega ^{/})W_{i}(\omega ^{/})\alpha _{i}(\omega )W_{j}^{\ast
}(\omega ^{/})\alpha _{j}^{\ast }(\omega ^{/})  \nonumber \\
&&+\omega \delta (\omega -\omega ^{/})\rho _{c}(\omega )[\pi ^{2}+|z(\omega
)|^{2}]W_{i}(\omega )\alpha _{i}(\omega )W_{j}^{\ast }(\omega )\alpha
_{j}^{\ast }(\omega )\}
\end{eqnarray}
The last term is just $\omega \delta (\omega -\omega ^{/})/\rho (\omega )$
using the normalization condition Eq.(\ref{anorm3}), the $\rho _{c}(\omega )$
factor cancelling out.. The next step is to eliminate the $F_{ij}$ using Eq.(%
\ref{z1}) for the $\alpha _{i}(\omega )$ twice. After further algebra using
Eq.(\ref{delta_tech2}) again we find that: 
\begin{eqnarray}
\int d\Delta \rho _{c}(\Delta )\Delta \beta (\Delta ,\omega )\beta ^{\ast
}(\Delta ,\omega ^{/}) &=&\omega \delta (\omega -\omega ^{/})/\rho (\omega
)+\sum_{i}\omega _{i}\alpha _{i}(\omega )\alpha _{i}^{\ast }(\omega ^{/}) 
\nonumber \\
&&-(\omega +\omega ^{/})\sum_{i}\alpha _{i}(\omega )\alpha _{i}^{\ast
}(\omega ^{/})+\sum_{ij(j\neq i)}\nu _{ji}^{\ast }\alpha _{i}(\omega )\alpha
_{j}^{\ast }(\omega ^{/})  \label{t2.eq}
\end{eqnarray}
The results in Eqs.(\ref{t2.eq}), (\ref{t4.eq}) and (\ref{t5.eq}) can be
substituted back into Eq.(\ref{Iww.eq}) for $I(\omega ,\omega ^{/})$. It is
found that there is extensive cancellation leading to the final expression: 
\begin{equation}
I(\omega ,\omega ^{/})=\omega \delta (\omega -\omega ^{/})/\rho (\omega )
\end{equation}
and hence the Hamiltonian $\hat{H}_{F}$ in Eq.(\ref{Hstart.eq}) is now in
its true mode form:

\begin{equation}
\hat{H}_{F}=\int d\omega \rho (\omega )\,\hbar \omega \hat{A}^{\dagger
}(\omega )\hat{A}(\omega )
\end{equation}
thus showing that the true and quasi mode forms of $\hat{H}_{F}$ are equal.

\bigskip

\label{References}

\clearpage

\clearpage

\end{document}